\documentclass[prl,twocolumn]{revtex4-1}

\usepackage{amsfonts,amsmath,amssymb,graphicx}
\usepackage[usenames,dvipsnames]{xcolor}
\usepackage{pdfpages}
\makeatletter
\AtBeginDocument{\let\LS@rot\@undefined}
\makeatother

\newcommand{\be}{\begin{eqnarray}}
\newcommand{\ee}{\end{eqnarray}}

\newcommand{\bra}[1]{\langle{#1}|}
\newcommand{\ket}[1]{|{#1}\rangle}

\newcommand{\catpm}{\ket{\mathcal{C}^\pm_\alpha}}

\newcommand{\Ep}{\mathcal{E}_\textnormal{p}}

\newcommand{\catpmi}{\ket{\mathcal{C}^\pm_{\alpha_0}}}
\newcommand{\catpi}{\ket{\mathcal{C}^+_{\alpha_0}}}
\newcommand{\catmi}{\ket{\mathcal{C}^-_{\alpha_0}}}
\newcommand{\hta}{\hat{a}}

\newcommand{\zero}{\bar{0}}
\newcommand{\one}{\bar{1}}

\newcommand{\Ez}{\mathcal{E}_\mathrm{z}}

\begin{document}

\title{Engineering the quantum states of light in a Kerr-nonlinear resonator by two-photon driving}
 
\author{Shruti Puri}
\affiliation{Institut quantique and D\'{e}partment de Physique, Universit\'{e} de Sherbrooke, Sherbrooke, Qu\'{e}bec, Canada J1K 2R1}
\author{Samuel Boutin}
\affiliation{Institut quantique and D\'{e}partment de Physique, Universit\'{e} de Sherbrooke, Sherbrooke, Qu\'{e}bec, Canada J1K 2R1}
\author{Alexandre Blais}
\affiliation{Institut quantique and D\'{e}partment de Physique, Universit\'{e} de Sherbrooke, Sherbrooke, Qu\'{e}bec, Canada J1K 2R1}
\affiliation{Canadian Institute for Advanced Research, Toronto, Canada}%



\begin{abstract}
{Photonic cat states stored in high-Q resonators show great promise for hardware efficient universal quantum computing. We propose an approach to efficiently prepare such cat states in a Kerr-nonlinear resonator by the use of a two-photon drive. Significantly, we show that this preparation is robust against single-photon loss. An outcome of this observation is that a two-photon drive can eliminate  undesirable phase evolution induced by a Kerr nonlinearity. By exploiting the concept of transitionless quantum driving, we moreover demonstrate how non-adiabatic initialization of cat states is possible. Finally, we present a universal set of quantum logical gates that can be performed on the engineered eigenspace of such a two-photon driven resonator and discuss a possible realization using superconducting circuits. The robustness of the engineered subspace to higher-order circuit nonlinearities makes this implementation favourable for scalable quantum computation.}

\end{abstract}

\maketitle
\section{Introduction}
Characterized by photon-photon interaction, Kerr-nonlinear resonators (KNR) display very rich physics and are consequently the focus of much theoretical and experimental work~\cite{dykman2012fluctuating}. These nonlinear oscillators exhibit bifurcation~\cite{siddiqi2005direct}, can be used to generate squeezed radiation and for quantum limited amplification~\cite{yurke1988observation,castellanos-beltran:2008a}, and have been proposed as a resource for quantum logic~\cite{munro:2005b}. Moreover, a KNR initialized in a coherent state evolves to a quantum superposition of out-of-phase coherent states, also known as a cat state~\cite{yurke:1986a}. In practice, Kerr nonlinearities $K$ in atomic systems are, however, often small in comparison to photon loss rate $\kappa$~\cite{boyd2003nonlinear}, making the observation of these non-classical states of light difficult. As an alternative approach, strong photon-photon interaction can readily be realized in superconducting quantum circuits, with $K/\kappa \sim 30$  demonstrated experimentally~\cite{kirchmair2013observation}. This has led to the observation of cat states in the transient dynamics of a KNR realized by coupling a superconducting qubit to a microwave resonator~\cite{kirchmair2013observation}. These photonic cat states play an important role in understanding the role of decoherence in macroscopic systems~\cite{zurek2003decoherence}, in precision measurements~\cite{munro2002weak} and are useful for quantum computation~\cite{ralph2003quantum,albert:2016a}. 
However, because of their sensitivity to undesirable interactions and photon loss, high-fidelity preparation and manipulation of these states is challenging. 

To address this problem, new ideas building on engineered dissipation and taking advantage of the strong nonlinearities that are possible with superconducting circuits have recently been explored theoretically and experimentally~\cite{leghtas2013deterministic,mirrahimi2014,vlastakis2013,leghtas2013hardware,wang2016schrodinger,ofek2016demonstrating}.
One such approach, known as the qcMAP gate, relies on the strong dispersive qubit-field interaction that is possible in circuit QED~\cite{gambetta:2006a} to transfer an arbitrary state of a superconducting qubit into a multi-legged cat state~\cite{leghtas2013deterministic,vlastakis2013,wang2016schrodinger}. This method is, however, susceptible to single-photon loss that decoheres the cat. This loss also reduces the amplitude of the cat, something that must be compensated for by re-pumping in order to avoid significant overlap between the coherent states~\cite{vlastakis2013,leghtas2013deterministic}. A second approach exploits engineered two-photon dissipation realized by coupling a superconducting qubit to two microwave cavities~\cite{mirrahimi2014,leghtas2015confining}. In the absence of single-photon loss, the steady-state of the field is a cat state whose parity depends on the initial number state of the field. To preserve coherence of the cat, an important experimental challenge is that the rate of single-photon loss must be much smaller than the rate of two-photon loss. 

In this paper we propose an experimentally simple alternative approach to encode and stabilize cat states based on two-photon driving of a KNR. This method takes advantage of the fact that the coherent states $\ket{\pm\alpha}$ and, consequently the cat states $\catpm=\mathcal{N}^\pm_\alpha(\ket{\alpha}\pm\ket{-\alpha})$ with $\mathcal{N}^\pm_\alpha=1/\sqrt{2(1\pm e^{-2|\alpha|^2})}$, are degenerate eigenstates of the KNR under two-photon driving.
Remarkably, this property holds true even in the presence of single-photon loss making this protocol particularly robust and obviating the need for energy re-pumping. Moreover, in contrast to the above-mentioned scheme, cat state preparation with this approach does not require dissipation but rather relies on adiabatically turning on the two-photon drive, the number state $\ket{0/1}$ evolving into $\ket{\mathcal{C}^{+/-}_{\alpha(t)}}$. We find that the fidelity of this  preparation approaches unity when the Kerr nonlinearity $K$ is large with respect to the photon loss rate $\kappa$, something that is easily realized in current circuit QED experiments. By exploiting the concept of transitionless quantum driving, we show that rapid, non-adiabatic cat state preparation is possible by controlling the amplitude and phase of the two-photon drive~\cite{berry2009transitionless}.

While large Kerr nonlinearities can be used to produce cat states, it also leads to undesired deformations of these states~\cite{yurke:1986a,kirchmair2013observation}. This deformation is problematic for qubit-based schemes because of the spurious Kerr nonlinearity inherited by the field from the qubit~\cite{maxime2,nigg:2012a}. This affects, for example, the qcMAP protocol where the qubit-induced Kerr nonlinearity leads to undesirable phase evolution and distortion of the cat state. Although this deterministic phase evolution can be corrected with qubit-induced-gates, this exposes the field to the decoherence channel of the qubit~\cite{heeres2015}. Moreover, in the presence of photons loss, this phase evolution leads to non-deterministic phase errors~\cite{ofek2016demonstrating}. We show how the addition of a two-photon drive of appropriate amplitude and phase during the qcMAP cancels this distortion and the corresponding dephasing. 

Taking advantage of the engineered subspace of a two-photon driven KNR, we consider a universal set of gates for an encoding where the coherent states $\{\ket{+\alpha},\ket{-\alpha}\}$ are mapped to the logical states $\{\ket{\bar{0}},\ket{\bar{1}}\}$. This mapping is possible because of the quasi-orthogonality of coherent states for large $\alpha$~\cite{mirrahimi2014}. We show that high-fidelity operations can be realized with realistic parameters. Finally, we discuss realizations based on superconducting Josephson parametric amplifiers which allow the implementation of a two-photon drive along with a Kerr nonlinearity. This simple setup is attractive for building a large scale quantum computing architecture.

\section{Results}
Our starting point is the two-photon driven KNR Hamiltonian in a frame rotating at the resonator frequency
\be
\hat{H}_0=-K\hat{a}^\dag \hat{a}^\dag \hat{a} \hat{a}+(\Ep \hat{a}^{\dag 2}+\Ep^* \hat{a}^2).
\label{eq:HCassinian}
\ee
In the above expression, $K$ is the amplitude of the Kerr nonlinearity and $\Ep$ the amplitude of the two-photon drive. The above Hamiltonian, known as the Cassinian oscillator Hamiltonian~\cite{wielinga1993quantum}, can be re-written as
\be
\hat{H}_0=-K\left(a^{\dag 2}-\frac{\Ep^*}{K}\right)\left(a^{2}-\frac{\Ep}{K}\right)+\frac{|\Ep|^2}{K}.
\ee
This form of the Hamiltonian illustrates that the two coherent states $\ket{\pm\alpha}$ with $\alpha=(\Ep/K)^{1/2}$, which are the eigenstates of the annihilation operator $\hta$, are also degenerate eigenstates of Eq.~\eqref{eq:HCassinian} with energy $|\Ep|^2/K$. Equivalently, the even-odd parity states $\catpm$ are also the eigenstates of $\hat{H}_0$. This argument can be generalized to Hamiltonians of the form $-K\hat{a}^{\dag n} \hat{a}^n+(\Ep \hat{a}^{\dag n}+\Ep^* \hat{a}^n)$ that have a set of $n$ coherent states as degenerate eigenstates (see Methods).

In the presence of single-photon loss, the resonator state evolves according to the master equation $\dot{\hat{\rho}}=-i(\hat{H}_\mathrm{eff}\hat{\rho}-\hat{\rho}\hat{H}_\mathrm{eff}^\dag)+\kappa \hta\hat{\rho} \hta^\dag$, with the non-Hermitian effective Hamiltonian $\hat{H}_\mathrm{eff}=\hat{H}_0-i\kappa \hta^\dag \hta/2$~\cite{walls2007quantum}. While the steady-state of this master equation can be obtained analytically~\cite{meaney2014quantum,minganti2016exact}, it is simple to show (see Methods) that for $\kappa/8|K\alpha_0^2|\ll 1$ 
the coherent states $\ket{\pm\alpha_0}=\ket{\pm r_0e^{i\theta_0}}$ are degenerate eigenstates of $\hat{H}_\mathrm{eff}$ with
\be
r_0=\left({\frac{{{4\mathcal{\Ep}}^2-\kappa^2/4}}{4K^2}}\right)^{1/4},
\;
\tan2\theta_0=\frac{\kappa}{\sqrt{16\Ep^2-\kappa^2}}.
\label{alphaSS}
\ee
This reduces to the eigenstates of $\hat{H}_0$ in the absence of photon loss. The angle $\theta_0$ is determined by $\Ep$, with $\theta_0<0$ ($\theta_0>0$) for $\Ep>0$ ($\Ep<0$). The last term of the master equation, $\kappa a\hat{\rho} a^\dag$, induces nondeterministic quantum jumps between the even and the odd parity cat states, $\catpi$ and $\catmi$, leading to decoherence, but not to leakage out of the degenerate subspace $\{\catpmi\}$. 
In steady-state, the density matrix therefore takes the form $\hat{\rho}_\mathrm{s}=(\ket{\alpha_0}\bra{\alpha_0}+\ket{-\alpha_0}\bra{-\alpha_0})/2$~\cite{meaney2014quantum,minganti2016exact} (see Methods). 
\begin{figure}
 \centering
 \includegraphics[width=0.70\columnwidth]{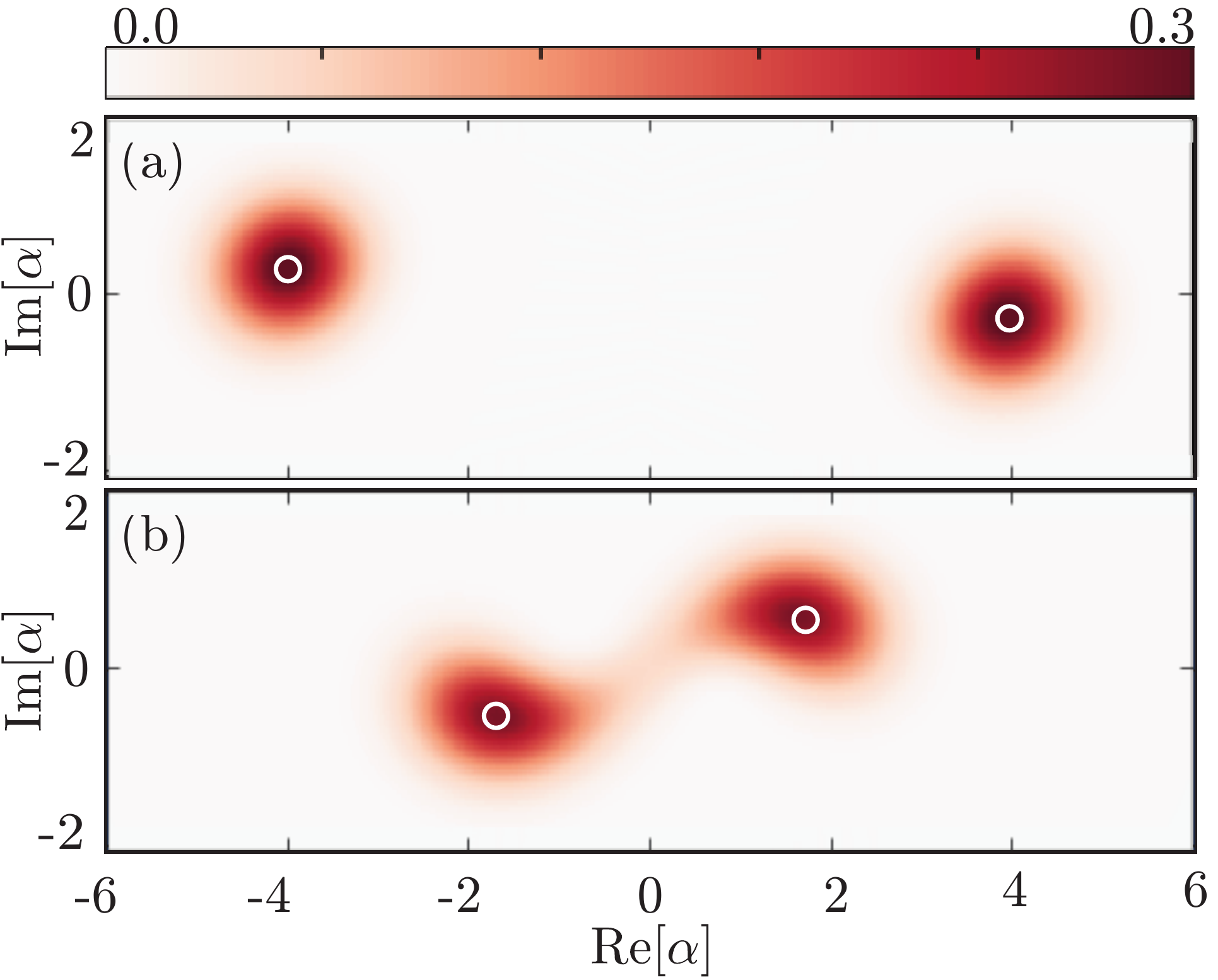}
 \caption{Steady-state Wigner function of a two-photon driven KNR with $|K|/\kappa=1/8$ and (a) $\Ep=16K$, $K>0$ and (b) $\Ep=4K$, $K<0$, corresponding to $\kappa/8|K\alpha_0^2|\sim1/16$ and $\sim1/4$ respectively. The white circles indicate the expected position of the coherent states following Eq.~\eqref{alphaSS}. \label{Wigsteadystate}}
 \end{figure}
 
Figure~\ref{Wigsteadystate} shows the steady-state Wigner function for $\kappa/8|K\alpha_0^2|\sim1/4$ and $\sim1/16$ obtained by numerical integration of the master equation~\cite{johansson2012qutip,johansson2013qutip}. Even for the relatively large value of $\kappa/8|K\alpha_0|^2\sim1/16$ shown in panel~a), the steady-state approaches the ideal case $\hat{\rho}_\mathrm{s}$ with a fidelity of $99.91\%$. As expected and evident from Fig.~\ref{Wigsteadystate}(b), the coherent states are deformed at the larger value of $\kappa/8|K\alpha_0|^2 \sim 1/4$ and the fidelity with respect to the ideal steady state is reduced to $96.55\%$. These numerical results confirm that, even in the presence of single-photon loss, it is possible to confine the state of the resonator to the manifold of coherent states $\ket{\pm\alpha_0}$. Although the photon loss channel remains the dominant source of error, the resonator can also have small amount of dephasing noise, which can cause jumps between $\ket{\alpha_0}$ and $\ket{-\alpha_0}$. With this bit-flip rate decreasing exponentially with $\alpha_0$~\cite{mirrahimi2014} (see also Supplementary Information S2), this channel is neglected here. 

\subsection{Adiabatic initialization of cat states:}
Going beyond steady-states, we now describe a protocol to deterministically prepare cat states. The vacuum $\ket{n=0}$ and the single-photon Fock state $\ket{n=1}$ are the two-degenerate eigenstates of the undriven KNR. Under the application of a time-dependent two-photon drive $\Ep(t)$, the instantaneous eigenstates of the system are the degenerate states $\ket{\pm\alpha_0(t)}$ (or equivalently $\ket{\mathcal{C}^\pm_{\alpha_0(t)}}$), where $\alpha_0(t)$ is given by Eq.~\eqref{alphaSS}. Since the two-photon drive preserves parity, under adiabatic increase of $\Ep(t)$, the vacuum state $\ket{0}$ evolves to the even parity cat state $\ket{\mathcal{C}^+_{\alpha_0(t)}}$ while the single-photon Fock state evolves to the odd parity cat state $\ket{\mathcal{C}^-_{\alpha_0(t)}}$ (see Supplementary Information S3 for the evolution of the energy spectrum). To demonstrate this deterministic preparation, we take as an example $\Ep(t)=\Ep^0[1-\exp(-t^4/\tau^4)]$ such that for $t\gg\tau$, $\Ep(t)\sim \Ep^0=4K$ with $\tau K=5$ to satisfy the adiabatic condition. Without photon loss, the fidelity of the resulting cat state at $t=6.5/K$ is 99.9$\%$ while for $K/\kappa=250$~\cite{bourassa2012josephson} the fidelity at $t=6.5/K$ is reduced to $98.3\%$. 


\subsection{High-fidelity nonadiabatic initialization:}
\begin{figure}
 \centering
 \includegraphics[width=0.85\columnwidth]{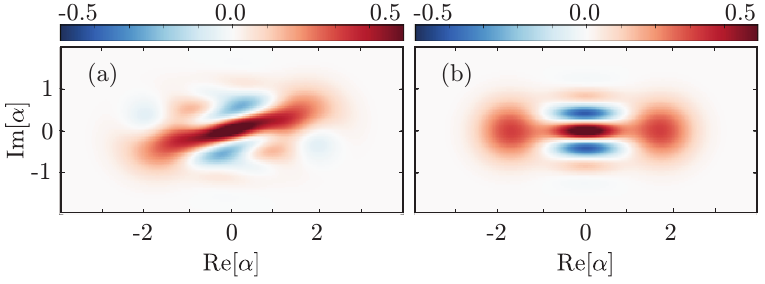}
 \caption{Wigner function for a KNR initialized in vacuum $\ket{0}$ and driven by (a) a single parametric drive $\Ep=\Ep^0[1-\exp(-t^4/\tau^4)]$ (b) with two orthogonal parametric drives, $\Ep=\Ep^0[1-\exp(-t^4/\tau^4)]$ and $\Ep'(t)=i{\dot{\alpha}_0(t)N^-_{\alpha_0(t)}}/(1+2\alpha_0(t))$, where $\alpha_0(t)=\sqrt{\Ep(t)/K}$. The Wigner function is plotted at time $t=1.37\tau$, with $\tau=1/K$, $\Ep^0=4K$. Without the auxiliary drive $\Ep'$ the non-adiabatic driving of the system results in an imperfect cat state. However, the auxiliary drive induces counter-adiabatic terms, resulting in near perfect initialization of the cat state. At $t=1.3\tau$, the fidelity with respect to $\ket{\mathcal{C}^+_2}$ is $99.9\%$ for $\kappa=0$ and $99.5\%$ for $K/\kappa=250$. \label{cat_init_tqd}}
 \end{figure}

To speed up the adiabatic preparation described above, we follow the approach of transitionless driving~\cite{berry2009transitionless, demirplak2003adiabatic, demirplak2005assisted}. This technique relies on introducing an auxiliary counter-adiabatic Hamiltonian, $\hat{H}'(t)=i[\ket{\dot{\psi}_n(t)}\bra{\psi_n(t)}-\ket{\psi_n(t)}\bra{\dot{\psi}_n(t)}]$, chosen such that the system follows the instantaneous eigenstate $\ket{\psi_n(t)}$ of the system Hamiltonian $\hat{H}_0(t)$ even under nonadiabatic changes of the system parameters. This idea has been experimentally demonstrated with Bose-Einstein condensates in optical lattices~\cite{bason2012high} and nitrogen vacancy centres in diamonds~\cite{zhang2013experimental}. Here, to prepare the even parity cat-state $\ket{\mathcal{C}^+_{\alpha_0(t)}}$, the required counter-adiabatic Hamiltonian is 
\begin{align}
\hat{H}'(t)&=i\frac{\dot{\alpha}_0(t)}{\mathcal{N}^-_{\alpha_0(t)}}\left[\hta^\dag\ket{\mathcal{C}^-_{\alpha_0(t)}}\bra{\mathcal{C}^+_{\alpha_0(t)}}-\ket{\mathcal{C}^+_{\alpha_0(t)}}\bra{\mathcal{C}^-_{\alpha_0(t)}}\hta\right]\label{tqd1}.
\end{align}
While exact, this does not correspond to an easily realizable Hamiltonian. It can, however, be approximated to (see Methods),
\begin{align}
\hat{H}'(t)&\sim i\frac{\dot{\alpha}_0(t)}{\mathcal{N}^-_{\alpha_0(t)}[1+2\alpha_0(t)]}(\hta^{\dag 2}-\hta^2)\label{tqd},
\end{align}
which can be implemented with an additional two-photon drive orthogonal to $\Ep(t)$. As an illustration of this method, we reconsider the example presented in the previous section now with the much shorter evolution time of $\tau=1/K$.  As shown by the Wigner function in Fig.~\ref{cat_init_tqd}(a), without the additional two-photon drive of Eq.~\eqref{tqd}, the state at time $t=1.37/K$ is highly distorted. On the other hand, and as illustrated in Fig.~\ref{cat_init_tqd}(b), initialization with the appropriate auxiliary orthogonal two-photon drive leads to cat-state fidelities of $99.9\%$ with $\kappa=0$ and $99.5\%$ with $\kappa=K/250$. In other words, we find that the protocol is made $\sim 5$ times faster by the addition of the orthogonal drive, thereby improving the fidelity in the presence of single-photon loss. These results, obtained with the analytical expression of Eq.~\eqref{tqd}, can be further improved upon using numerical optimal control~\cite{khaneja:2005a}. For example, using the approach recently described in Ref.~\cite{boutin2016resonator}, we find that cat states can be initialized in times as short as $0.3/K$ with fidelity $99.995\%$ (see Supplementary Information S4). Adiabatic cat state preparation with two-photon driving was also investigated in a noiseless idealized KNR~\cite{goto:2016b,goto:2016a}. These previous studies lack the crucial examination of eigenspace distortion that arise, as will be discussed below, during gate operations and  fall short of accounting for higher-order nonlinearities that exist in realistic physical implementations.

\subsection{Realization with superconducting circuits:} One standard approach to realize a two-photon driven Kerr-nonlinear resonator is to terminate a $\lambda/4$ microwave resonator with a flux-pumped SQUID, a device known as a Josephson parametric amplifier~\cite{yamamoto:2008a,wustmann2013parametric,Krantz:2016aa} (see also Supplementary Information S5). The non-linear inductance of the SQUID induces a Kerr nonlinearity and a two-photon drive is introduced by the modulation of the flux-pump at twice the resonator frequency.  As an illustrative example, with a realistic JPA Kerr-nonlinearity of $K/2\pi=750$ KHz it is possible to encode a cat state with $\alpha_0=2$ in a time $0.3/K=63.6$ ns using the transitionless driving approach with numerically optimized pulse shape. We have, moreover, simulated the cat state initialization protocol under the exact Hamiltonian of a JPA including the full Josephson junction cosine potential. As discussed in the Supplementary Information S5, the results are essentially unchanged showing that the strong state confinement to the coherent states $\ket{\pm\alpha_0}$ is also robust against higher-order nonlinearities that will arise in a circuit implementation of these ideas. An alternative realization of the two-photon driven KNR is based on a 3D microwave cavity coupled to a Josephson junction. The non-linear inductance of the junction induces a Kerr nonlinearity, while a microwave drive on the junction at the 3D cavity frequency introduces the required two-photon drive~\cite{mirrahimi2014,leghtas2015confining}.


We note that the engineered dissipation approach of Refs.~\cite{mirrahimi2014,leghtas2015confining} also relies on a two-photon drive to achieve confinement to the subspace of two coherent states with opposite phases. There, the two-photon drives is used to induce two-photon loss at a rate $\kappa_\mathrm{2ph}$. This rate must be made large with respect to the single-photon loss rate $\kappa$ for high fidelity initialization of cat states, something which is challenging experimentally. In contrast, the present approach does not rely on dissipation but rather takes advantage of the large Kerr-nonlinearity $K$ that is easily realized in superconducting quantum circuits. Even in the presence of two-photon loss, robust confinement is obtained if $K>\kappa_\mathrm{2ph}$, a condition that is easily satisfied in practice.



\subsection{Stabilization of cat states against Kerr induced rotation and dephasing:}
\begin{figure}
\centering
\includegraphics[width=0.9\columnwidth]{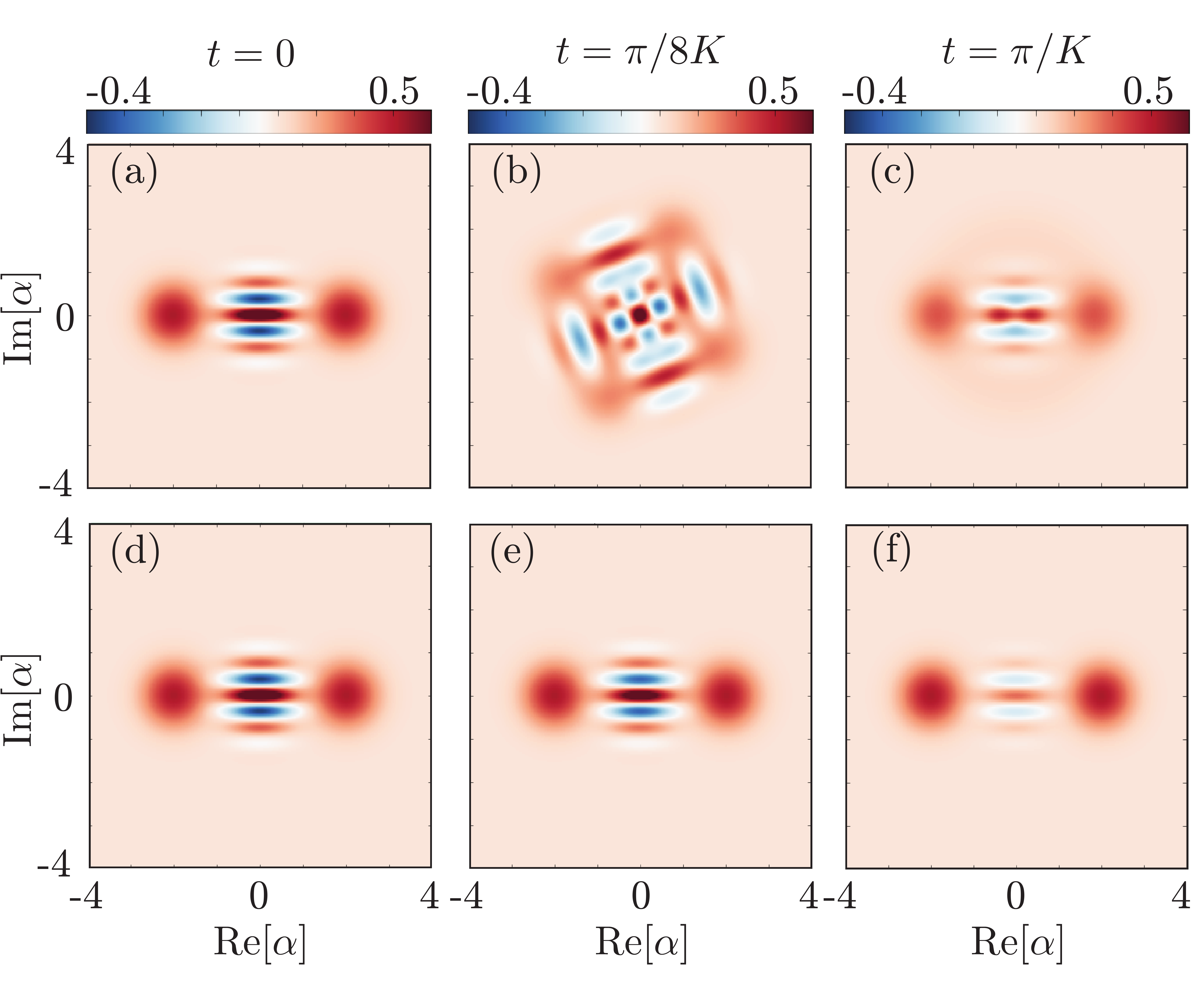}
 \caption{Wigner functions at different times for a lossy KNR initialized to $\ket{\mathcal{C}^+_2}$ without (a-c) and with (d-f) two-photon driving. $K/\kappa=20$ and $\Ep\sim 4K$. \label{CatStab}}

 \end{figure}

Even with high-fidelity cat state preparation, it is important to limit the unwanted phase evolution and dephasing arising from Kerr nonlinearity and single-photon loss. We now illustrate, with two examples, how a two-photon drive of appropriate amplitude and phase can correct this unwanted evolution. First consider a resonator deterministically initialized to $\ket{\mathcal{C}_\alpha^+}$. Figure~\ref{CatStab}(a-c) illustrates the evolution of this initial state in the absence of two-photon drive. Kerr nonlinearity leads to deterministic deformation of the state~\cite{yurke:1986a,kirchmair2013observation} which, in the presence of single-photon loss, also induces additional dephasing. This results in a reduction of the contrast of the Wigner function fringes, a reduction of the separation of the cat components and a broadening of these components. As a result, the fidelity of $\ket{\mathcal{C}_\alpha^\pm}$ decreases faster in a KNR than in a linear resonator (see Supplementary Information S9). While the deterministic phase rotation can be accounted for and corrected in a simple way, this is not the case for Kerr-induced dephasing~\cite{heeres2015}. Fig.~\ref{CatStab}(d-f) illustrates the same initial cat state now stabilized against Kerr-induced rotation and dephasing by the application of a two-photon drive. This drive is chosen such that its amplitude $\Ep$ satisfies Eq.~\eqref{alphaSS}. The confinement in phase space provided by the two-photon driven KNR prevents amplitude damping of the stabilized coherent states $\ket{\pm\alpha_0}$. As a result, the cat state fidelity in this system decreases more slowly in time that in a linear resonator. As a simple extension, we also find that it is possible to stabilize coherent states against Kerr-induced rotation and dephasing (see Supplementary Information S1). These somewhat counterintuitive results shows that, even in the presence of loss, a Gaussian drive (i.e. two-photon drive) can completely remove the highly non-Gaussian effect of a Kerr nonlinearity.

\begin{figure}
 \centering
 \includegraphics[width=0.60\columnwidth]{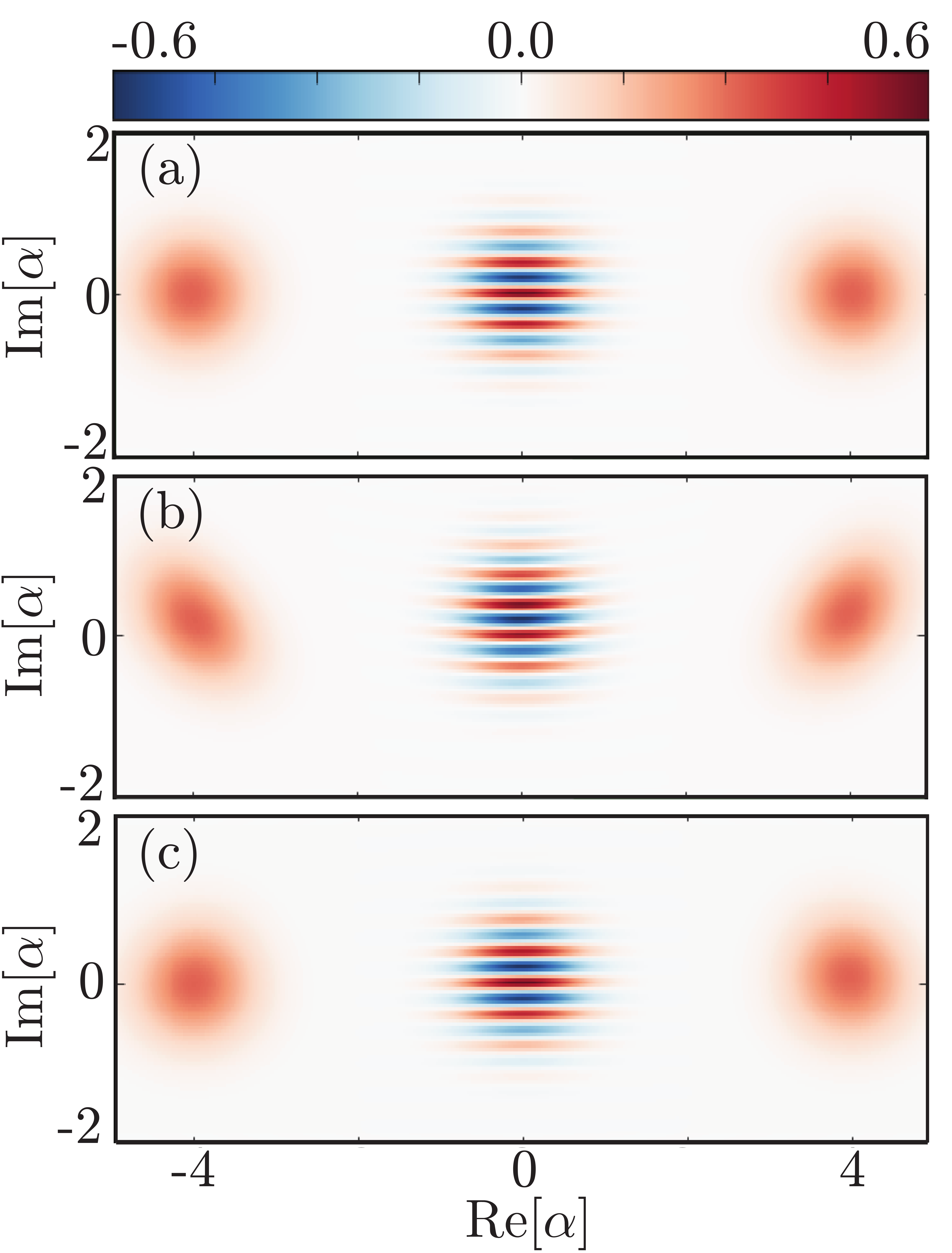}
 \caption{Wigner function of final state under qcMAP gate with (a) ideal dispersive Hamiltonian, (b) full Jaynes-Cummings Hamiltonian and (c) full Jaynes-Cummings Hamiltonian and two-photon drive.  \label{qcmap_cat}}
 \end{figure}

As a second example, we consider the qcMAP gate for cat state preparation, a protocol that relies on the strong dispersive qubit-resonator interaction that is realized in circuit QED~\cite{leghtas2013deterministic}. In practice, this strong interaction is accompanied by a qubit-induced Kerr nonlinearity of the field~\cite{maxime2,nigg:2012a}. As a result, even at modest $\alpha$, cat states suffer from deformations~\cite{vlastakis2013}. This effect is illustrated in Fig.~\ref{qcmap_cat}(a,b) which shows the cat state obtained from qcMAP under ideal dispersive interaction (ignoring any Kerr nonlinearities) and under the full Jaynes-Cummings Hamiltonian, respectively. Distortions are apparent in panel b) and the fidelity to the ideal cat is reduced to $94.1\%$. In contrast, Fig.~\ref{qcmap_cat}(c) shows the same Wigner function prepared using the qcMAP protocol with the full Jaynes-Cummings interaction and an additional two-photon drive. The resulting fidelity is $99.4\%$, approaching the fidelity of 99.8\% obtained under the ideal, but not realistic, dispersive Hamiltonian. The amplitude of the two-photon drive was optimized numerically to take into account the qubit-induced Kerr nonlinearity (see Supplementary Information S10).

\subsection{Universal quantum logic gates:}
Following the general approach of Ref.~\cite{mirrahimi2014}, we now turn to the realization of a universal set of gates in the two-photon driven KNR. Taking advantage of the quasi-orthogonality of coherent states for large $\alpha$, both the $\{\catpmi\}$ and the $\{\ket{\pm\alpha_0}\}$ basis can be used as logical states. Here, we choose the latter which we will now refer to as $\{\ket{\bar{0}},\ket{\bar{1}}\}$. With this choice, a logical 
Z rotation can be realized by lifting the degeneracy between $\ket{\bar{0}}$ and $\ket{\bar{1}}$ using a single-photon drive in combination to $\hat H_0$: $
\hat{H}_\mathrm{z}=\hat{H}_0+\mathcal{E}_\mathrm{z}(\hat  a^\dag+
\hat a)$. For $|\mathcal{E}_\mathrm{z}|\ll |4K\alpha_0^3|$ and $\Ep$ real, the only effect of this additional drive is to lift the degeneracy by $\delta_z=4\mathcal{E}_\mathrm{z}\alpha_0$ (Supplementary Information S6). Indeed, in the space spanned by $\{\ket{\bar{0}},\ket{\bar{1}}\}$, the single-photon drive Hamiltonian can be expressed as 
$\bar I\mathcal{E}_\mathrm{z}(\hat a^\dag + \hat a)\bar I=\delta_z\bar{\sigma}_\mathrm{z}/2$, 
where $\bar I = \ket\zero\bra\zero+\ket\one\bra\one$ and $\bar{\sigma}_\mathrm{z}=\ket{\zero}\bra{\zero}-\ket{\one}\bra{\one}$. Numerical simulations of this process for a time $\tau=1/\delta_z$, corresponding to the gate $\hat{R}_\mathrm{\bar z}(\pi)$, with the resonator initialized to $\catpi$ and the choices $\Ep=4K$, $\Ez=0.8K$ leads to a fidelity of 99.9$\%$ with $\kappa = 0$ and $99.5\%$ for $K/\kappa=250$. 
Increasing $\Ez$, so that the condition $|\mathcal{E}_\mathrm{z}|\ll |4K\alpha_0^3|$ is no longer satisfied, distorts the eigenstates and as a consequence the fidelity of the gate decreases. The dependence of the gate fidelity on the strength of the single photon drive is examined further in Supplementary Information S8. A similar scheme for single-qubit rotation has been proposed for resonators with engineered two-photon loss~\cite{mirrahimi2014}. However, this requires the drive strength to be significantly smaller than the two-photon loss rate which is typically of the order of $50-100$ kHz~\cite{leghtas2015confining}, thereby leading to long gate times. 

The strong state confinement resulting from the two-photon driven KNR prevents  population  transfer between the two logical states, making it difficult to implement $X$ rotations. One approach to implement $\hat{R}_\mathrm{\bar x}(\pi/2)$ is to temporarily remove the two-photon drive and let the state evolve under the Kerr Hamiltonian~\cite{mirrahimi2014}. Alternatively, an arbitrary $\hat{R}_\mathrm{\bar x}(\theta)$ can be realized by introducing a detuning between the two-photon drive and the resonator corresponding to the Hamiltonian $\hat{H}_\mathrm{x}=\hat{H}_0+\delta_x\hta^\dag\hta$. For $\delta_x\ll2\Ep$ (Supplementary Information S7), this can be understood by projecting the number operator in the logical basis: $\bar I\hta^\dag\hta \bar I = |\alpha_0|^2\bar{I}-|\alpha_0|^2e^{-2|\alpha_0|^2}\bar\sigma_x$. Despite the exponential reduction with $\alpha_0$ of the effective Rabi frequency, high-fidelity rotations can be achieved. Numerical simulations on a resonator initialized to $\ket\zero$  and for a time $\tau=\pi/(4\delta_x|\alpha_0|^2e^{-2|\alpha_0|^2})$, corresponding to the gate $\hat{R}_\mathrm{\bar x}(\pi/2)$, leads to a fidelity of $99.7\%$ for $\kappa=0$ and $98.6\%$ for $K/\kappa=250$ with $\Ep=K$ and $\delta_x=K/3$. Similarly to the $Z$ rotations, the fidelity of the $X$ gate also decreases if the condition $\delta_x\ll2\Ep$ is not met (see Supplementary Information S8).

To complete the set of universal gates, an entangling gate between the field stored in two distinct resonators, or alternatively two modes of a single resonator, is needed. 
From the discussion on the $\hat{R}_\mathrm{\bar z}(\theta)$ gate, it follows that a $\bar\sigma_\mathrm{z1}\bar\sigma_\mathrm{z2}$ interaction between the two fields is obtained by linearly coupling the two-photon driven KNRs, the Hamiltonian now reading $\hat{H}_\mathrm{zz}=\hat{H}_{01}+\hat{H}_{02}+\mathcal{E}_\mathrm{zz}(\hta_1^\dag\hta_2 + \hta_1\hta_2^\dag)$. To simplify the discussion, the two resonators are assumed to be identical with $\hat{H}_{0i}=-K\hta_i^{\dag 2}\hta_i^2+\Ep(\hta_i^{\dag 2}+\hta_i^2)$. Expressed in the logical basis, the bilinear coupling Hamiltonian takes the desired form $\delta_{zz}\bar\sigma_\mathrm{z1}\bar\sigma_\mathrm{z2}$, with $\delta_{zz}=4\mathcal{E}_\mathrm{zz}|\alpha_0|^2$.
In order to demonstrate this gate, we simulate the master equation under $\hat{H}_\mathrm{zz}$ with the resonators initialized to the product state $\ket{\mathcal{C}^+_{\alpha_0}}\otimes \ket{\mathcal{C}^+_{\alpha_0}}$ and $\Ep=4K$, $\mathcal{E}_\mathrm{zz}=K/5$. As expected, the initial product state is transformed to the maximally entangled state $(\ket{\zero,\zero}+i\ket{\zero,\one}+i\ket{\one,\zero}+\ket{\one,\one})/2$ at $t=\pi/2\delta_{zz}$ with fidelity $F=99.99\%$ for $\kappa=0$ and $F=94\%$ for $K/\kappa=250$. Supplementary Information S8 examines the fidelity dependence on the strength of the two-photon drive. Similar approaches for $Z$ rotations and entangling gate have been presented before~\cite{goto:2016a}, however without the crucial analysis of the restrictions on the amplitude of the single-photon drive and strength of the single-photon exchange coupling.

\section{Discussion}

To summarize, we have shown that, in the presence of a two-photon drive, the eigenspace of a KNR can be engineered to be two out-of-phase coherent states that are robust against single-photon loss. This quantum state engineering offers a practical way to correct the undesirable effects of Kerr nonlinearity in applications such as the qcMAP gate. We have also described protocols for fast-high fidelity initialization and manipulation cat states for quantum information processing. This approach offers significant improvements over previous techniques based on dispersive qubit-resonator interactions or reservoir engineering. These results suggest a minimal approach to prepare and manipulate cat states of the field of a microwave resonator using only a Josephson parametric amplifier and are of immediate practical importance for realization of a scalable, hardware efficient platform for quantum computation. Furthermore, the observation that $n$ coherent states are the degenerate eigenstates of the Hamiltonian $\hat{H}=-K\hta^{\dag n}\hta^n+\Ep(a^{\dag n}+a^n)$ provides an approach for initializing $n$-component cat states. Such a Hamiltonian could be implemented with a JPA, in which the cosine potential of a Josephson junction supplies the required nonlinearity and flux modulation through the SQUID loop at $n-$times the resonator frequency triggers the $n$-photon drive. Our work opens new directions for the JPA as a powerful device for implementing quantum algorithms based on multi-component cats.

\section{Methods}
\subsection{Eigenstates of the $n$-photon driven Hamiltonian:}
Consider the Hamiltonian 
\be
\begin{split}
\hat{H}_n
&=-K\hta^{\dag n}\hta^n+(\Ep \hta^{\dag n} + \Ep^* \hta^{n})\\
&=-K\left(\hta^{\dag n}-\frac{\Ep^*}{K}\right)\left(\hta^{n}-\frac{\Ep}{K}\right)+\frac{|\Ep|^2}{K}.
\end{split}
\ee
The second form makes it clear that the coherent state $\ket{\alpha}$ with $\alpha^n-\Ep/K=0$ is an eigenstate of $\hat{H}_n$. Thus, in general, there are $n$ coherent states that are the degenerate eigenstates of $\hat{H}_n$ with energy $|\Ep|^2/K$. 
\subsection{Effective Hamiltonian and steady-state:}

Under single-photon loss, the system's master equation takes the form~\cite{walls2007quantum}
\begin{equation}\label{eq:ME}
\dot{\hat{\rho}}=-i(\hat{H}_\mathrm{eff}\hat{\rho}-\hat{\rho}\hat{H}_\mathrm{eff}^\dag)+\kappa \hta\hat{\rho} \hta^\dag,
\end{equation}
where $\hat{H}_\mathrm{eff} = \hat{H}_0-i\kappa \hta^\dag \hta/2$ and $\hat{H}_0=-K\hat{a}^\dag \hat{a}^\dag \hat{a} \hat{a}+(\Ep \hat{a}^{\dag 2} + \Ep^* \hat{a}^2)$. Under displacement transformation $D(\alpha_0)=\exp(\alpha_0\hta^\dag-\alpha_0\hta)$, $\hat{H}_\mathrm{eff}$ reads
\begin{equation}
\begin{split}
\hat{H}'_\mathrm{eff}
& =D^\dag(\alpha_0)\hat{H}_\mathrm{eff}D(\alpha_0)\\
& = \left[(-2K\alpha_0^2\alpha_0^*+2\Ep\alpha_0^*-i\frac{\kappa}{2}\alpha_0)\hta^\dag+\mathrm{h.c.}\right]\\
&+\left[(-K\alpha_0^2+\Ep)\hta^{\dag 2}+\mathrm{h.c.}\right]-4K|\alpha|^2\hta^\dag \hta\\
&-i\frac{\kappa}{2}\hta^\dag \hta-K\hta^{\dag 2}\hta^2-(2K\alpha_0 \hta^{\dag 2}a+\mathrm{h.c.}),
\end{split}
\end{equation}
where we have dropped the constant term $E=-K|\alpha_0|^4+\Ep^*\alpha_0^2+\Ep\alpha_0^{*2}-i\kappa|\alpha_0|^2/2$ that represents a shift in energy of the non-Hermitan effective Hamiltonian. We take $\alpha_0$ to satisfy 
\be
-2K\alpha_0^2\alpha_0^*+2\Ep\alpha_0^*-i\frac{\kappa}{2}\alpha_0=0,
\label{C1}
\ee
such as to cancel the first line of $\hat{H}'_\mathrm{eff}$ which now reads
\be
\begin{split}
\hat{H}'_\mathrm{eff}
&=
[(-K\alpha_0^2+\Ep)\hta^{\dag 2}+\mathrm{h.c.}]-(4K|\alpha_0|^2+i\frac{\kappa}{2})\hta^\dag \hta\\
&-K\hta^{\dag 2}\hta^2-2K\alpha_0 \hta^{\dag 2}a-2K\alpha_0^* \hta^\dag \hta^2.
\label{Heff}
\end{split}
\ee
Eq.~\eqref{C1} is satisfied for $\alpha_0=0,\,\pm r_0e^{i\theta_0}$ where
\be
\begin{split}
r_0&=\left({\frac{{{4\mathcal{\Ep}}^2-\kappa^2/4}}{4K^2}}\right)^{1/4},\\
\quad \theta_0&=\frac{1}{2}\tan^{-1}\left(\frac{\kappa}{\sqrt{16\Ep^2-\kappa^2}}\right).
\end{split}
\ee

For $\alpha_0=0$, the first two terms of Eq.~\eqref{Heff} represent a near resonant parametric drive of strength $\Ep$. This results in large fluctuations making the system unstable around $\alpha_0=0$. On the other hand, for $\alpha_0=\pm r_0e^{i\theta_0}$, the displaced effective Hamiltonian can be rewritten as
\be
\begin{split}
\hat{H}'_\mathrm{eff}&=\frac{1}{2}\left[i\frac{\kappa\alpha_0}{2\alpha_0^*}\hta^{\dag 2}+\mathrm{c.c.}\right]
-(4K|\alpha_0|^2+i\frac{\kappa}{2})\hta^\dag \hta\\
&-K\hta^{\dag 2}\hta^2-2K\alpha_0 \hta^{\dag 2}a-2K\alpha_0^* \hta^\dag \hta^2.
\end{split}
\label{Heff2}
\ee
The first two terms of Eq.~\eqref{Heff2} now represent a parametric drive whose amplitude has an absolute value of $\kappa/2$ and is detuned by $4K|\alpha_0|^2 +i\kappa/2 \approx4K|\alpha_0|^2$. In other words, the effect of single-photon loss $\kappa$ is to squeeze the field around $\alpha_0=\pm r_0e^{i\theta_0}$ leading to increased quantum fluctuations. For $\kappa\ll 8K|\alpha_0|^2$, the resulting fluctuations are, however, small and $\ket{0}$ remains an eigenstate in the displaced frame. This implies that, back in the lab frame, $\ket{\pm\alpha_0}$ are the degenerate eigenstates of $\hat{H}_\mathrm{eff}$. As a result, $\hat{\rho}_{\mathrm{s}}=(\ket{\alpha_0}\bra{\alpha_0}+\ket{-\alpha_0}\bra{-\alpha_0})/2$ is a steady-state of Eq.~\eqref{eq:ME}. It is, moreover, the unique steady-state of this system since only the two eigenstates $\ket{\pm\alpha_0}$ of the effective Hamiltonian are also invariant under the quantum jump operator $\hta$~\cite{kraus2008preparation}. Following the analysis here, it is also possible to characterize the effect of, for example, single-photon drive, detuning, etc (see Supplementary Notes).

\subsection{Cat state decoherence under single-photon loss:} In the previous section, we saw that the coherent states $\ket{\pm\alpha_0}$ are eigenstates of the two-photon driven KNR even in the presence of single-photon loss. However, this loss channel results in decoherence of superpositions of these two states, i.e.~of cat states. Indeed, the last term of the master equation Eq.~\eqref{eq:ME}, $\kappa a\hat{\rho} a^\dag$, transforms the even parity cat state $\catpi$ to the odd parity cat state $\catmi$ and vice-versa. This results in decoherence and reduction in the contrast of the Wigner function fringes. The rate of this phase decay is given by $\gamma=\kappa |\alpha_0-(-\alpha_0)|^2/2=2\kappa|\alpha_0|^2$. 

Consider for example the cat state initialization protocol with $\Ep=\Ep^0[1-\exp(-t^4/\tau^4)]$ and $\Ep^0=4K$ so that $\alpha_0(t)=2\sqrt{[1-\exp(-t^4/\tau^4)]}$. The phase error during this  initialization can be estimated to be $\exp(-2\int\kappa|\alpha_0(t)|^2\mathrm{d}t)=0.016$, resulting in a fidelity of $98.4\%$. This estimate compares very well with the numerically estimated fidelity quoted earlier in the manuscript ($98.3\%$). 

\subsection{Additional Hamiltonian for faster than adiabatic initialization of cat state}
Consider the exact Hamiltonian in Eq.~\eqref{tqd1} required for transitionless quantum driving. At short times $t\sim0$, we have that $\alpha_0(t)\sim0$ and as a result $\ket{\mathcal{C}^+_{0}}\sim \ket{n=0}$ and $\ket{\mathcal{C}^-_{0}}\sim \ket{n=1}$.  Therefore, $\left[\hta^\dag\ket{\mathcal{C}^-_{\alpha_0(t)}}\bra{\mathcal{C}^+_{\alpha_0(t)}}-\ket{\mathcal{C}^+_{\alpha_0(t)}}\bra{\mathcal{C}^-_{\alpha_0(t)}}\hta\right]\sim[\hta^\dag\ket{1}\bra{0}-\ket{0}\bra{1}\hta]\sim \hta^{\dag 2}-\hta^2$. On the contrary, at long time the coherent states become quasi-orthogonal and a single photon jump leads to the transition between even and odd photon number cat states. This suggests that if $\alpha_0(t)\gg 1$, it is possible to approximate 
$\left[\hta^\dag\ket{\mathcal{C}^-_{\alpha_0(t)}}\bra{\mathcal{C}^+_{\alpha_0(t)}}-\ket{\mathcal{C}^+_{\alpha_0(t)}}\bra{\mathcal{C}^-_{\alpha_0(t)}}\hta\right]\sim (\hta^{\dag 2}-\hta^2)/2\alpha_0(t)$ in the restricted coherent state basis. Therefore, in order to reconcile both short and long time behaviour, we choose, $\left[\hta^\dag\ket{\mathcal{C}^-_{\alpha_0(t)}}\bra{\mathcal{C}^+_{\alpha_0(t)}}-\ket{\mathcal{C}^+_{\alpha_0(t)}}\bra{\mathcal{C}^-_{\alpha_0(t)}}\hta\right]\sim (\hta^{\dag 2}-\hta^2)/[1+2\alpha_0(t)]$ to obtain Eq.~\eqref{tqd}.

\section*{Acknowledgements}
We thank M.~Mirrahimi, A.~Grimsmo, and C.~Andersen for useful discussions. \\
\section*{Competing Interests} 
The authors declare that they have no competing interests.\\
\section*{Contributions}
S. P. and A. B. conceived and developed the idea, S. B. contributed to the circuit analysis and developed the code based on GRAPE for pulse optimization. \\
\section*{Funding}
This work was supported by the Army Research Office under Grant W911NF-14-1-0078 and by NSERC. This research was undertaken thanks in part to funding from the Canada First Research Excellence Fund.
\section*{References}
\bibliography{Manuscript_Arxiv.bbl}{}

\begin{thebibliography}{10}
\expandafter\ifx\csname url\endcsname\relax
  \def\url#1{\texttt{#1}}\fi
\expandafter\ifx\csname urlprefix\endcsname\relax\def\urlprefix{URL }\fi
\providecommand{\bibinfo}[2]{#2}
\providecommand{\eprint}[2][]{\url{#2}}

\bibitem{dykman2012fluctuating}
\bibinfo{author}{Dykman, M.}
\newblock \emph{\bibinfo{title}{Fluctuating nonlinear oscillators: from
  nanomechanics to quantum superconducting circuits}} (\bibinfo{publisher}{OUP
  Oxford}, \bibinfo{year}{2012}).

\bibitem{siddiqi2005direct}
\bibinfo{author}{Siddiqi, I.} \emph{et~al.}
\newblock \bibinfo{title}{Direct observation of dynamical bifurcation between
  two driven oscillation states of a josephson junction}.
\newblock \emph{\bibinfo{journal}{Physical review letters}}
  \textbf{\bibinfo{volume}{94}}, \bibinfo{pages}{027005}
  (\bibinfo{year}{2005}).

\bibitem{yurke1988observation}
\bibinfo{author}{Yurke, B.} \emph{et~al.}
\newblock \bibinfo{title}{Observation of 4.2-k equilibrium-noise squeezing via
  a josephson-parametric amplifier}.
\newblock \emph{\bibinfo{journal}{Physical Review Letters}}
  \textbf{\bibinfo{volume}{60}}, \bibinfo{pages}{764} (\bibinfo{year}{1988}).

\bibitem{castellanos-beltran:2008a}
\bibinfo{author}{Castellanos-Beltran, M.~A.}, \bibinfo{author}{Irwin, K.~D.},
  \bibinfo{author}{Hilton, G.~C.}, \bibinfo{author}{Vale, L.~R.} \&
  \bibinfo{author}{Lehnert, K.~W.}
\newblock \bibinfo{title}{Amplification and squeezing of quantum noise with a
  tunable josephson metamaterial}.
\newblock \emph{\bibinfo{journal}{Nat Phys}} \textbf{\bibinfo{volume}{4}},
  \bibinfo{pages}{929--931} (\bibinfo{year}{2008}).
\newblock \urlprefix\url{http://dx.doi.org/10.1038/nphys1090}.

\bibitem{munro:2005b}
\bibinfo{author}{Munro, W.~J.}, \bibinfo{author}{Nemoto, K.} \&
  \bibinfo{author}{Spiller, T.~P.}
\newblock \bibinfo{title}{Weak nonlinearities: a new route to optical quantum
  computation}.
\newblock \emph{\bibinfo{journal}{New Journal of Physics}}
  \textbf{\bibinfo{volume}{7}}, \bibinfo{pages}{137} (\bibinfo{year}{2005}).
\newblock \urlprefix\url{http://stacks.iop.org/1367-2630/7/i=1/a=137}.

\bibitem{yurke:1986a}
\bibinfo{author}{Yurke, B.} \& \bibinfo{author}{Stoler, D.}
\newblock \bibinfo{title}{Generating quantum mechanical superpositions of
  macroscopically distinguishable states via amplitude dispersion}.
\newblock \emph{\bibinfo{journal}{Phys. Rev. Lett.}}
  \textbf{\bibinfo{volume}{57}}, \bibinfo{pages}{13--16}
  (\bibinfo{year}{1986}).
\newblock \urlprefix\url{http://link.aps.org/doi/10.1103/PhysRevLett.57.13}.

\bibitem{boyd2003nonlinear}
\bibinfo{author}{Boyd, R.~W.}
\newblock \emph{\bibinfo{title}{Nonlinear optics}}
  (\bibinfo{publisher}{Academic press}, \bibinfo{year}{2003}).

\bibitem{kirchmair2013observation}
\bibinfo{author}{Kirchmair, G.} \emph{et~al.}
\newblock \bibinfo{title}{Observation of quantum state collapse and revival due
  to the single-photon kerr effect}.
\newblock \emph{\bibinfo{journal}{Nature}} \textbf{\bibinfo{volume}{495}},
  \bibinfo{pages}{205--209} (\bibinfo{year}{2013}).

\bibitem{zurek2003decoherence}
\bibinfo{author}{Zurek, W.~H.}
\newblock \bibinfo{title}{Decoherence, einselection, and the quantum origins of
  the classical}.
\newblock \emph{\bibinfo{journal}{Reviews of Modern Physics}}
  \textbf{\bibinfo{volume}{75}}, \bibinfo{pages}{715} (\bibinfo{year}{2003}).

\bibitem{munro2002weak}
\bibinfo{author}{Munro, W.~J.}, \bibinfo{author}{Nemoto, K.},
  \bibinfo{author}{Milburn, G.~J.} \& \bibinfo{author}{Braunstein, S.~L.}
\newblock \bibinfo{title}{Weak-force detection with superposed coherent
  states}.
\newblock \emph{\bibinfo{journal}{Physical Review A}}
  \textbf{\bibinfo{volume}{66}}, \bibinfo{pages}{023819}
  (\bibinfo{year}{2002}).

\bibitem{ralph2003quantum}
\bibinfo{author}{Ralph, T.}, \bibinfo{author}{Gilchrist, A.},
  \bibinfo{author}{Milburn, G.~J.}, \bibinfo{author}{Munro, W.~J.} \&
  \bibinfo{author}{Glancy, S.}
\newblock \bibinfo{title}{Quantum computation with optical coherent states}.
\newblock \emph{\bibinfo{journal}{Physical Review A}}
  \textbf{\bibinfo{volume}{68}}, \bibinfo{pages}{042319}
  (\bibinfo{year}{2003}).

\bibitem{albert:2016a}
\bibinfo{author}{Albert, V.~V.} \emph{et~al.}
\newblock \bibinfo{title}{Holonomic quantum control with continuous variable
  systems}.
\newblock \emph{\bibinfo{journal}{Phys. Rev. Lett.}}
  \textbf{\bibinfo{volume}{116}}, \bibinfo{pages}{140502}
  (\bibinfo{year}{2016}).
\newblock
  \urlprefix\url{http://link.aps.org/doi/10.1103/PhysRevLett.116.140502}.

\bibitem{leghtas2013deterministic}
\bibinfo{author}{Leghtas, Z.} \emph{et~al.}
\newblock \bibinfo{title}{Deterministic protocol for mapping a qubit to
  coherent state superpositions in a cavity}.
\newblock \emph{\bibinfo{journal}{Physical Review A}}
  \textbf{\bibinfo{volume}{87}}, \bibinfo{pages}{042315}
  (\bibinfo{year}{2013}).

\bibitem{mirrahimi2014}
\bibinfo{author}{Mirrahimi, M.} \emph{et~al.}
\newblock \bibinfo{title}{Dynamically protected cat-qubits: a new paradigm for
  universal quantum computation}.
\newblock \emph{\bibinfo{journal}{New Journal of Physics}}
  \textbf{\bibinfo{volume}{16}}, \bibinfo{pages}{045014}
  (\bibinfo{year}{2014}).

\bibitem{vlastakis2013}
\bibinfo{author}{Vlastakis, B.} \emph{et~al.}
\newblock \bibinfo{title}{Deterministically encoding quantum information using
  100-photon schr{\"o}dinger cat states}.
\newblock \emph{\bibinfo{journal}{Science}} \textbf{\bibinfo{volume}{342}},
  \bibinfo{pages}{607--610} (\bibinfo{year}{2013}).

\bibitem{leghtas2013hardware}
\bibinfo{author}{Leghtas, Z.} \emph{et~al.}
\newblock \bibinfo{title}{Hardware-efficient autonomous quantum memory
  protection}.
\newblock \emph{\bibinfo{journal}{Physical Review Letters}}
  \textbf{\bibinfo{volume}{111}}, \bibinfo{pages}{120501}
  (\bibinfo{year}{2013}).

\bibitem{wang2016schrodinger}
\bibinfo{author}{Wang, C.} \emph{et~al.}
\newblock \bibinfo{title}{A schrodinger cat living in two boxes}.
\newblock \emph{\bibinfo{journal}{arXiv preprint arXiv:1601.05505}}
  (\bibinfo{year}{2016}).

\bibitem{ofek2016demonstrating}
\bibinfo{author}{Ofek, N.} \emph{et~al.}
\newblock \bibinfo{title}{Demonstrating quantum error correction that extends
  the lifetime of quantum information}.
\newblock \emph{\bibinfo{journal}{arXiv preprint arXiv:1602.04768}}
  (\bibinfo{year}{2016}).

\bibitem{gambetta:2006a}
\bibinfo{author}{Gambetta, J.} \emph{et~al.}
\newblock \bibinfo{title}{Qubit-photon interactions in a cavity:
  Measurement-induced dephasing and number splitting}.
\newblock \emph{\bibinfo{journal}{Physical Review A (Atomic, Molecular, and
  Optical Physics)}} \textbf{\bibinfo{volume}{74}}, \bibinfo{pages}{042318}
  (\bibinfo{year}{2006}).
\newblock \urlprefix\url{http://link.aps.org/abstract/PRA/v74/e042318}.

\bibitem{leghtas2015confining}
\bibinfo{author}{Leghtas, Z.} \emph{et~al.}
\newblock \bibinfo{title}{Confining the state of light to a quantum manifold by
  engineered two-photon loss}.
\newblock \emph{\bibinfo{journal}{Science}} \textbf{\bibinfo{volume}{347}},
  \bibinfo{pages}{853--857} (\bibinfo{year}{2015}).

\bibitem{berry2009transitionless}
\bibinfo{author}{Berry, M.}
\newblock \bibinfo{title}{Transitionless quantum driving}.
\newblock \emph{\bibinfo{journal}{Journal of Physics A: Mathematical and
  Theoretical}} \textbf{\bibinfo{volume}{42}}, \bibinfo{pages}{365303}
  (\bibinfo{year}{2009}).

\bibitem{maxime2}
\bibinfo{author}{Boissonneault, M.}, \bibinfo{author}{Gambetta, J.~M.} \&
  \bibinfo{author}{Blais, A.}
\newblock \bibinfo{title}{Dispersive regime of circuit qed: Photon-dependent
  qubit dephasing and relaxation rates}.
\newblock \emph{\bibinfo{journal}{Physical Review A}}
  \textbf{\bibinfo{volume}{79}}, \bibinfo{pages}{013819}
  (\bibinfo{year}{2009}).

\bibitem{nigg:2012a}
\bibinfo{author}{Nigg, S.~E.} \emph{et~al.}
\newblock \bibinfo{title}{Black-box superconducting circuit quantization}.
\newblock \emph{\bibinfo{journal}{Phys. Rev. Lett.}}
  \textbf{\bibinfo{volume}{108}}, \bibinfo{pages}{240502}
  (\bibinfo{year}{2012}).
\newblock
  \urlprefix\url{http://link.aps.org/doi/10.1103/PhysRevLett.108.240502}.

\bibitem{heeres2015}
\bibinfo{author}{Heeres, R.~W.} \emph{et~al.}
\newblock \bibinfo{title}{Cavity state manipulation using photon-number
  selective phase gates}.
\newblock \emph{\bibinfo{journal}{arXiv preprint arXiv:1503.01496}}
  (\bibinfo{year}{2015}).

\bibitem{wielinga1993quantum}
\bibinfo{author}{Wielinga, B.} \& \bibinfo{author}{Milburn, G.}
\newblock \bibinfo{title}{Quantum tunneling in a kerr medium with parametric
  pumping}.
\newblock \emph{\bibinfo{journal}{Physical Review A}}
  \textbf{\bibinfo{volume}{48}}, \bibinfo{pages}{2494} (\bibinfo{year}{1993}).

\bibitem{walls2007quantum}
\bibinfo{author}{Walls, D.~F.} \& \bibinfo{author}{Milburn, G.~J.}
\newblock \emph{\bibinfo{title}{Quantum optics}} (\bibinfo{publisher}{Springer
  Science \& Business Media}, \bibinfo{year}{2007}).

\bibitem{meaney2014quantum}
\bibinfo{author}{Meaney, C.~H.}, \bibinfo{author}{Nha, H.},
  \bibinfo{author}{Duty, T.} \& \bibinfo{author}{Milburn, G.~J.}
\newblock \bibinfo{title}{Quantum and classical nonlinear dynamics in a
  microwave cavity}.
\newblock \emph{\bibinfo{journal}{EPJ Quantum Technology}}
  \textbf{\bibinfo{volume}{1}}, \bibinfo{pages}{1--23} (\bibinfo{year}{2014}).

\bibitem{minganti2016exact}
\bibinfo{author}{Minganti, F.}, \bibinfo{author}{Bartolo, N.},
  \bibinfo{author}{Lolli, J.}, \bibinfo{author}{Casteels, W.} \&
  \bibinfo{author}{Ciuti, C.}
\newblock \bibinfo{title}{Exact results for schr{\"o}dinger cats in
  driven-dissipative systems and their feedback control}.
\newblock \emph{\bibinfo{journal}{Scientific reports}}
  \textbf{\bibinfo{volume}{6}} (\bibinfo{year}{2016}).

\bibitem{johansson2012qutip}
\bibinfo{author}{Johansson, J.}, \bibinfo{author}{Nation, P.} \&
  \bibinfo{author}{Nori, F.}
\newblock \bibinfo{title}{Qutip: An open-source python framework for the
  dynamics of open quantum systems}.
\newblock \emph{\bibinfo{journal}{Computer Physics Communications}}
  \textbf{\bibinfo{volume}{183}}, \bibinfo{pages}{1760--1772}
  (\bibinfo{year}{2012}).

\bibitem{johansson2013qutip}
\bibinfo{author}{Johansson, J.}, \bibinfo{author}{Nation, P.} \&
  \bibinfo{author}{Nori, F.}
\newblock \bibinfo{title}{Qutip 2: A python framework for the dynamics of open
  quantum systems}.
\newblock \emph{\bibinfo{journal}{Computer Physics Communications}}
  \textbf{\bibinfo{volume}{184}}, \bibinfo{pages}{1234--1240}
  (\bibinfo{year}{2013}).

\bibitem{bourassa2012josephson}
\bibinfo{author}{Bourassa, J.}, \bibinfo{author}{Beaudoin, F.},
  \bibinfo{author}{Gambetta, J.~M.} \& \bibinfo{author}{Blais, A.}
\newblock \bibinfo{title}{Josephson-junction-embedded transmission-line
  resonators: From kerr medium to in-line transmon}.
\newblock \emph{\bibinfo{journal}{Physical Review A}}
  \textbf{\bibinfo{volume}{86}}, \bibinfo{pages}{013814}
  (\bibinfo{year}{2012}).

\bibitem{demirplak2003adiabatic}
\bibinfo{author}{Demirplak, M.} \& \bibinfo{author}{Rice, S.~A.}
\newblock \bibinfo{title}{Adiabatic population transfer with control fields}.
\newblock \emph{\bibinfo{journal}{The Journal of Physical Chemistry A}}
  \textbf{\bibinfo{volume}{107}}, \bibinfo{pages}{9937--9945}
  (\bibinfo{year}{2003}).

\bibitem{demirplak2005assisted}
\bibinfo{author}{Demirplak, M.} \& \bibinfo{author}{Rice, S.~A.}
\newblock \bibinfo{title}{Assisted adiabatic passage revisited}.
\newblock \emph{\bibinfo{journal}{The Journal of Physical Chemistry B}}
  \textbf{\bibinfo{volume}{109}}, \bibinfo{pages}{6838--6844}
  (\bibinfo{year}{2005}).

\bibitem{bason2012high}
\bibinfo{author}{Bason, M.~G.} \emph{et~al.}
\newblock \bibinfo{title}{High-fidelity quantum driving}.
\newblock \emph{\bibinfo{journal}{Nature Physics}}
  \textbf{\bibinfo{volume}{8}}, \bibinfo{pages}{147--152}
  (\bibinfo{year}{2012}).

\bibitem{zhang2013experimental}
\bibinfo{author}{Zhang, J.} \emph{et~al.}
\newblock \bibinfo{title}{Experimental implementation of assisted quantum
  adiabatic passage in a single spin}.
\newblock \emph{\bibinfo{journal}{Physical review letters}}
  \textbf{\bibinfo{volume}{110}}, \bibinfo{pages}{240501}
  (\bibinfo{year}{2013}).

\bibitem{khaneja:2005a}
\bibinfo{author}{Khaneja, N.}, \bibinfo{author}{Reiss, T.},
  \bibinfo{author}{Kehlet, C.}, \bibinfo{author}{Schulte-Herbruggen, T.} \&
  \bibinfo{author}{Glaser, S.~J.}
\newblock \bibinfo{title}{Optimal control of coupled spin dynamics: design of
  nmr pulse sequences by gradient ascent algorithms}.
\newblock \emph{\bibinfo{journal}{Journal of Magnetic Resonance}}
  \textbf{\bibinfo{volume}{172}}, \bibinfo{pages}{296--305}
  (\bibinfo{year}{2005}).

\bibitem{boutin2016resonator}
\bibinfo{author}{Boutin, S.}, \bibinfo{author}{Andersen, C.~K.},
  \bibinfo{author}{Venkatraman, J.}, \bibinfo{author}{Ferris, A.~J.} \&
  \bibinfo{author}{Blais, A.}
\newblock \bibinfo{title}{Resonator reset in circuit qed by optimal control for
  large open quantum systems}.
\newblock \emph{\bibinfo{journal}{arXiv preprint arXiv:1609.03170}}
  (\bibinfo{year}{2016}).

\bibitem{goto:2016b}
\bibinfo{author}{Goto, H.}
\newblock \bibinfo{title}{Bifurcation-based adiabatic quantum computation with
  a nonlinear oscillator network}.
\newblock \emph{\bibinfo{journal}{Scientific Reports}}
  \textbf{\bibinfo{volume}{6}}, \bibinfo{pages}{21686 EP --}
  (\bibinfo{year}{2016}).
\newblock \urlprefix\url{http://dx.doi.org/10.1038/srep21686}.

\bibitem{goto:2016a}
\bibinfo{author}{Goto, H.}
\newblock \bibinfo{title}{Universal quantum computation with a nonlinear
  oscillator network}.
\newblock \emph{\bibinfo{journal}{Phys. Rev. A}} \textbf{\bibinfo{volume}{93}},
  \bibinfo{pages}{050301} (\bibinfo{year}{2016}).
\newblock \urlprefix\url{http://link.aps.org/doi/10.1103/PhysRevA.93.050301}.

\bibitem{yamamoto:2008a}
\bibinfo{author}{Yamamoto, T.} \emph{et~al.}
\newblock \bibinfo{title}{Flux-driven josephson parametric amplifier}.
\newblock \emph{\bibinfo{journal}{Applied Physics Letters}}
  \textbf{\bibinfo{volume}{93}}, \bibinfo{pages}{042510}
  (\bibinfo{year}{2008}).
\newblock \urlprefix\url{http://link.aip.org/link/?APL/93/042510/1}.

\bibitem{wustmann2013parametric}
\bibinfo{author}{Wustmann, W.} \& \bibinfo{author}{Shumeiko, V.}
\newblock \bibinfo{title}{Parametric resonance in tunable superconducting
  cavities}.
\newblock \emph{\bibinfo{journal}{Physical Review B}}
  \textbf{\bibinfo{volume}{87}}, \bibinfo{pages}{184501}
  (\bibinfo{year}{2013}).

\bibitem{Krantz:2016aa}
\bibinfo{author}{Krantz, P.} \emph{et~al.}
\newblock \bibinfo{title}{Single-shot read-out of a superconducting qubit using
  a josephson parametric oscillator}.
\newblock \emph{\bibinfo{journal}{Nat Commun}} \textbf{\bibinfo{volume}{7}}
  (\bibinfo{year}{2016}).
\newblock \urlprefix\url{http://dx.doi.org/10.1038/ncomms11417}.

\bibitem{kraus2008preparation}
\bibinfo{author}{Kraus, B.} \emph{et~al.}
\newblock \bibinfo{title}{Preparation of entangled states by quantum markov
  processes}.
\newblock \emph{\bibinfo{journal}{Physical Review A}}
  \textbf{\bibinfo{volume}{78}}, \bibinfo{pages}{042307}
  (\bibinfo{year}{2008}).

\end{thebibliography}

\clearpage


\section*{Supplementary material}

# Supplementary Information: Engineering the quantum states of light in a Kerr-nonlinear resonator by two-photon driving

Shruti Puri[1], Samuel Boutin[1] and Alexandre Blais[1,2]

1. Institut quantique and Département de Physique,
Université de Sherbrooke, Sherbrooke, Québec, Canada J1K 2R1 and
2. Canadian Institute for Advanced Research, Toronto, Canada

## I. STABILIZATION OF COHERENT STATES IN A TWO-PHOTON DRIVEN KNR

In the manuscript (section Methods) it was shown that $\hat{\rho}_s = (|\alpha_0\rangle\langle\alpha_0| + |-\alpha_0\rangle\langle-\alpha_0|)/2$ is the unique steady-state of a two-photon driven KNR. It is worth pointing out that, although this is the unique steady-state, the time for the system to reach $\hat{\rho}_s$ can approach infinity as $|\alpha_0|$ increases because $\langle-\alpha_0|\alpha_0\rangle = \exp(-2|\alpha_0|^2) \sim 0$. As a result, a system initialized in the state $a|\alpha_0\rangle + b|-\alpha_0\rangle$ will evolve to $|a|^2|\alpha_0\rangle\langle\alpha_0| + |b|^2|-\alpha_0\rangle\langle-\alpha_0|$ only after a long time $t \gg 1/\kappa$ if $\alpha_0$ is large.

As a corollary to the above discussion, we find that the two-photon driven KNR initialized to either of the coherent states $|\pm\alpha_0\rangle$, remains in that state even in the presence of single-photon loss, as long as $\alpha_0$ satisfies Eq. (11) in the main text and $\kappa \ll 8K|\alpha_0|^2$. This is illustrated in Fig. S1, which shows the Wigner function obtained by numerical integration of the master equation for the system initialized to the coherent state $|\alpha_0\rangle$ without [Fig. S1(a-c)] and with [Fig. S1(d-f)] two-photon drive. We note that all simulations in this work were carried out with a standard master equation solver [1, 2] and with Hilbert space size large enough to ensure negligible truncation errors. For example, simulations with $|\alpha| = 1, 2, 4$ were carried out with a Hilbert space size of $N = 20, 40, 80$ respectively.

## II. EFFECT OF PHOTON DEPHASING AND TWO-PHOTON LOSS

In the main paper, we have not taken into account resonator dephasing and two-photon loss as they are typically negligible compared to single-photon loss. However, to complete the analysis we briefly discuss their effects here.

### A. Photon-dephasing

To take into account dephasing at a rate $\kappa_\phi$, the master equation takes the form

$$\dot{\rho} = -i(\hat{H}_{\text{eff}}\hat{\rho} - \hat{\rho}\hat{H}_{\text{eff}}^\dagger) + \kappa_\phi\hat{a}^\dagger\hat{a}\hat{\rho}\hat{a}^\dagger\hat{a}, \tag{S1}$$

with $\hat{H}_{\text{eff}} = \hat{H}_0 - i\kappa_\phi\hat{a}^\dagger\hat{a}\hat{a}^\dagger\hat{a}/2 = \hat{H}_0 - i\kappa_\phi\hat{a}^{\dagger 2}\hat{a}^2/2 - i\kappa_\phi\hat{a}^\dagger\hat{a}/2$. This effective Hamiltonian is equivalent to the effective Hamiltonan when $\kappa_\phi = 0$ but with a Kerr nonlinearity $K - i\kappa_\phi/2$ and single-photon loss $\kappa_\phi$. Following the derivation in the second section we find that the amplitude $\alpha_0$ must now satisfy

$$(-2K - i\kappa_\phi)\,\alpha_0^2\alpha_0^* + 2\mathcal{E}_p\alpha_0^* - i\frac{\kappa_\phi}{2}\alpha_0 = 0. \tag{S2}$$

For $\kappa_\phi \ll 4K|\alpha_0|^2$, the coherent states $|\pm\alpha_0\rangle$ are again the degenerate eigenstates of the system.

The action of the third term in the above master equation, $\hat{a}^\dagger\hat{a}\hat{\rho}\hat{a}^\dagger\hat{a}$, it to flip the state of the resonator from $|\alpha_0\rangle$ to $|-\alpha_0\rangle$ and vice-versa at a rate of $\kappa_\phi|\alpha_0|^2e^{-2|\alpha_0|^2}$.

### B. Two-photon loss

Two-photon loss, the process in which the system looses pairs of photons to the bath, often accompanies non-linear interactions [3]. However, ordinarily, the rate of two-photon dissipation is small. The master equation in the presence of such a loss (with rate $\kappa_{2ph}$) takes the form

$$\dot{\rho} = -i(\hat{H}_{\text{eff}}\hat{\rho} - \hat{\rho}\hat{H}_{\text{eff}}^\dagger) + \kappa_{2ph}\hat{a}^2\hat{\rho}\hat{a}^{\dagger 2}, \tag{S3}$$

where $\hat{H}_{\text{eff}} = \hat{H}_0 - i\kappa_{2ph}\hat{a}^{\dagger 2}\hat{a}^2/2$. This expression implies that two-photon loss effectively acts as a Kerr-nonlinearity of amplitude $-i\kappa_{2ph}/2$. As a result, the degenerate eigenstates of the effective Hamiltonian become $|\pm\alpha\rangle$ with

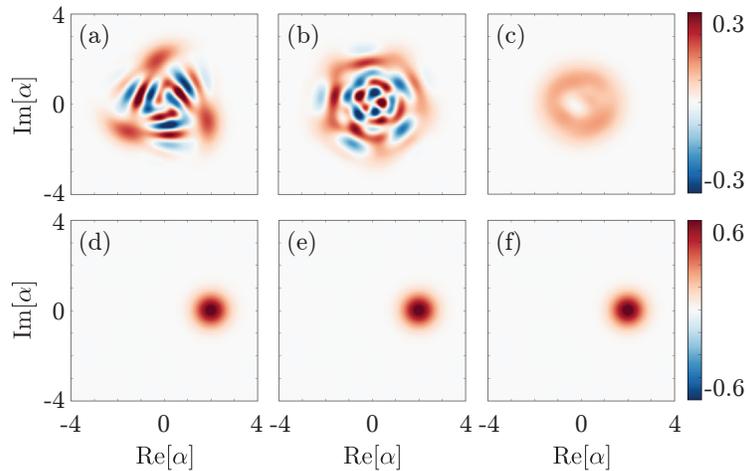

FIG. S1. Time evolution of the Wigner function when the resonator is initialized to the coherent state $|2\rangle$ and $\kappa = K/250$. In the absence of a parametric drive (a-c), the coherent state evolves under the Kerr Hamiltonian and will finally decay to the vacuum state $|0\rangle$. With a parametric drive $\mathcal{E}_\mathrm{p} \sim 4K$ (d-e) (satisfying Eq. (11) in the main text), the initial state is the eigenstate of the effective Hamiltonian and it remains in that state.

$\alpha = \sqrt{\mathcal{E}_\mathrm{p}/(K + i\kappa_{2ph}/2)}$. The two-photon jump operator, given by the second term in the above expression, does not cause a phase-flip or spin-flip since $\hat{a}^2|\pm\alpha\rangle = \alpha^2|\pm\alpha\rangle$ and $\hat{a}^2|\mathcal{C}_\alpha^\pm\rangle = \alpha^2|\mathcal{C}_\alpha^\pm\rangle$. As a result, two-photon loss is, by itself, not detrimental to cat state preparation. However, this deterministic phase rotation, in addition to non-deterministic single-photon loss can lead to additional dephasing. With typical parameter such as the one described in the main body of the paper and in section V of the supplement, $\kappa_{2ph} \ll K$ and dephasing is therefore negligible.

## III. EIGENSTATES AND EIGENVALUES OF THE TWO-PHOTON DRIVEN KNR

Here we numerically evaluate the eigenstates and eigenenergies of the time-dependent KNR driven by a two-photon process

$$\hat{H}_0(t) = K\hat{a}^\dagger\hat{a}^\dagger\hat{a}\hat{a} - [\mathcal{E}_\mathrm{p}(t)\hat{a}^{\dagger 2} + \mathcal{E}_\mathrm{p}(t)^*\hat{a}^2], \tag{S4}$$

with $\mathcal{E}_\mathrm{p}(t) = \mathcal{E}_\mathrm{p}^0[1 - \exp(-t^4/\tau^4)]$ so that $\mathcal{E}_\mathrm{p}(t = 0) = 0$, $\mathcal{E}_\mathrm{p}(t) \sim \mathcal{E}_\mathrm{p}^0 = 4K$ for $t \gg \tau$ and $\tau K = 5$. This particular time dependance was chosen to assure adiabaticity of the evolution. Other choices are possible and pulse shaping techniques could lead to fidelity increases. We also note that the sign of the Kerr nonlinearity as changed with respect to the main body of the paper.

We take $K > 0$ for which the ground states at $t = 0$ are the Fock states $|0\rangle, |1\rangle$ with energy $E_0 = E_1 = 0$. Under adiabatic evolution, these degenerate ground states transform to the instantaneous eigenstates of the Hamiltonian, $|\mathcal{C}_{\alpha_0(t)}^\pm\rangle$ for $t > \tau$ with energy $E_0 = E_1 = -\mathcal{E}_\mathrm{p}\alpha_0(t)^2$ and $\alpha_0 = \sqrt{\mathcal{E}_\mathrm{p}(t)/K}$. The eigenenergies of the ground, first and second instantaneous eigenstates are plotted in Fig. S2(a). The figure also shows the simulated Wigner functions corresponding to the instantaneous eigenstates at different times. Note that if $K < 0$ the states $|0\rangle, |1\rangle$ are excited states of the initial undriven Kerr Hamiltonian. However as the two-photon drive amplitude increases these states are slowly transformed to the eigenstates $|\mathcal{C}_{\alpha_0(t)}^\pm\rangle$, which are eigenstates but not necessarily the ground states of $\hat{H}_0(t)$. Figure S2(b) illustrates the time dependence of the eigenenergies and eigenstates when $K < 0$ and $\mathcal{E}_\mathrm{p}^0 = 4K$.

The eigenenergy spectrum illustrates that for adiabatic initialization of cat states in time $\tau$, $\Delta_{\min}\tau \gg 1$. Here $\Delta_{\min}$ is the minimum energy gap and, from Fig. S2, $\Delta_{\min} = 2K$. It is possible to increase this gap and therefore speed-up the initialization by introducing a time-dependent detuning $\delta(t)$ between the two-photon drive and the bare resonator frequency. The initialization protocol is then carried out by increasing the two-photon drive strength and decreasing the detuning from $\delta(0) = \delta_0$ to $\delta(\tau) = 0$. The detuning given by the Hamiltonian $\delta(t)\hat{a}^\dagger\hat{a}$ conserves parity at all times (that is, it does not mix the even and odd parity cat states) and increases the minimum energy gap during the adiabatic evolution, leading to a faster initialization. Consider for example a resonator subjected to the time-dependent Hamiltonian, $\hat{H}(t) = -K\hat{a}^\dagger\hat{a}^\dagger\hat{a}\hat{a} + (t/\tau)\mathcal{E}_\mathrm{p}^0[\hat{a}^{\dagger 2} + \hat{a}^2] - \delta_0(1 - t/\tau)\hat{a}^\dagger\hat{a}$, with $\mathcal{E}_\mathrm{p}^0 = 4K, \delta_0 = 1.7K$ and $\tau = 2K$. The minimum energy gap during this evolution is $\Delta_{\min} = 4.3K$. At $t = 0$, the Fock states are the eigenstates, whereas at $t = \tau$ the eigenstates are the cat states $|\mathcal{C}_{\alpha_0(t)}^\pm\rangle$. We find that such a resonator initialized to

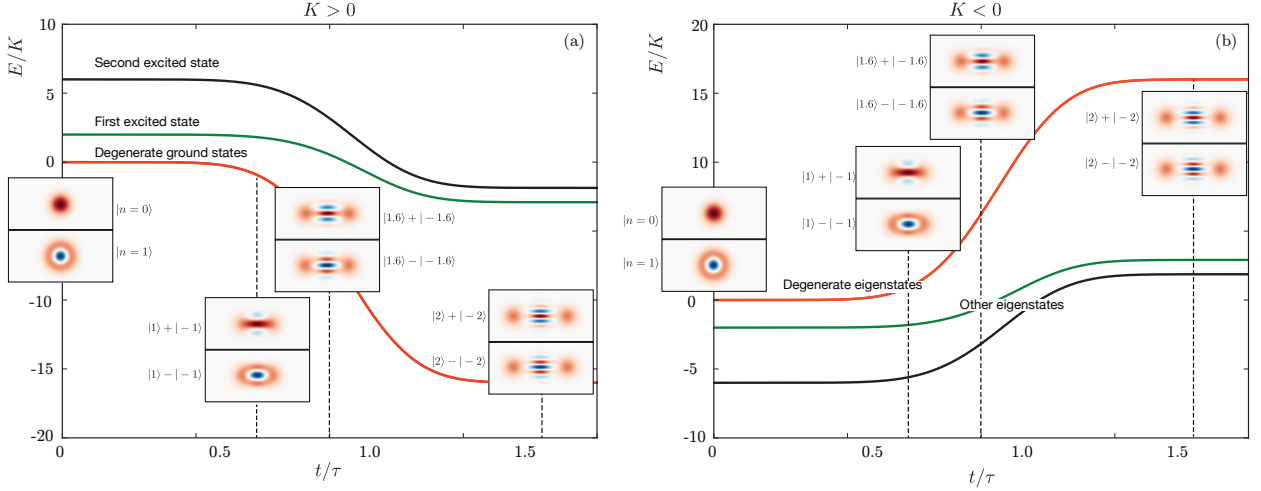

FIG. S2. Numerically evaluated eigenenergies and Wigner function of the first four eigenstates for (a) $K > 0$ (b) $K < 0$, $\mathcal{E}_\mathrm{p}(t) = \mathcal{E}_\mathrm{p}^0[1 - \exp(-t^4/\tau^4)]$, $\mathcal{E}_\mathrm{p}^0 = 4K$ and $\tau = 5K$.

vacuum evolves to the cat state $|\mathcal{C}_{\alpha_0(t)}^+\rangle$ at $t = \tau$ with a fidelity of 99.9% for $\kappa = 0$, and 99.3% for $\kappa = K/250$. If, on the other hand, $\delta_0 = 0$ and the initialization is carried out in a time $\tau = 2K$, then the cat state fidelity is reduced to 85.6% when $\kappa = 0$ and 84.9% when $\kappa = K/250$ because of non-adiabatic errors. As already mentioned, further speed-ups could be obtained by pulse shaping techniques.

## IV. PULSE OPTIMIZATION WITH GRAPE

An implementation of the Gradient Ascent Pulse Engineering (GRAPE) algorithm [4] was used to design the pulse for fast cat state initialization using the non-adiabatic protocol described in the main text. Following the result of the main text, fast initialization is achieved by evolution under the time dependent Hamiltonian $\hat{H}(t) = \hat{H}_0(t) + \hat{H}'(t)$, with

$$\hat{H}_0(t) = -K\hat{a}^{\dagger 2}\hat{a}^2 + \mathcal{E}_\mathrm{p,x}(t)(\hat{a}^{\dagger 2} + \hat{a}^2), \quad \hat{H}'(t) = i\mathcal{E}_\mathrm{p,y}(t)(\hat{a}^{\dagger 2} - \hat{a}^2).$$

(S5)

Our GRAPE implementation allows the restriction of the two-photon drive $\mathcal{E}_\mathrm{p,y}$ to zero at the beginning $t = 0$ and end $t = T$ of the protocol. The two-photon drive $\mathcal{E}_\mathrm{p,x}$ is restricted to $4K$ at $t = T$ to realize a stabilized cat $\mathcal{C}_2^+$ at the end of the protocol. Furthermore, in order to allow only realistic drive amplitudes during the evolution, the pulse amplitude is restricted such that $|\mathcal{E}_\mathrm{p,y}|, |\mathcal{E}_\mathrm{p,x}| < 6K$. The resulting pulse, optimized to yield the cat state $\mathcal{C}_2^+$ at $t = T = 0.3/K$ is shown in Fig. S3. The fidelity of the resulting cat state is 99.95% and the time step is chosen so that the time-scale for the modulation of the drive amplitude is realistic ($\geq 1$ ns).

## V. IMPLEMENTATION OF THE ENCODING SCHEME IN CQED WITH A JOSEPHSON PARAMETRIC AMPLIFIER

In this section, we present numerical simulations of the cat state preparation protocol with a Josephson parametric amplifier (JPA). The Hamiltonian of a lumped element JPA is given by [5–7]

$$\hat{H}(t) = \frac{\hat{q}^2}{2C} - 2E_\mathrm{J}\cos\left(\frac{\Phi(t)}{\phi_0}\right)\cos\left(\frac{\hat{\phi}}{\phi_0}\right),$$

(S6)

where

$$\hat{q} = i\sqrt{\frac{\hbar C\omega_\mathrm{r}}{2}}(\hat{a}^\dagger - \hat{a}), \quad \hat{\phi} = \sqrt{\frac{\hbar}{2C\omega_\mathrm{r}}}(\hat{a}^\dagger + \hat{a})$$

(S7)

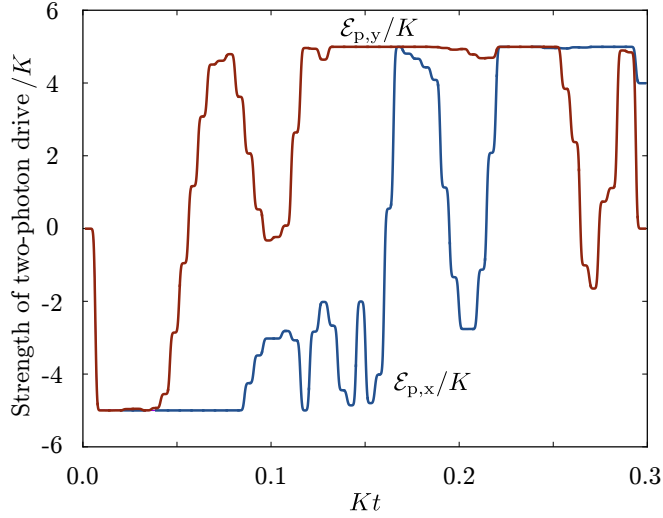

FIG. S3. Optimized pulse shapes using an implementation of the GRAPE algorithm [4] for $\mathcal{E}_{p,x}(t)/K$ (blue) and $\mathcal{E}_{p,y}(t)/K$ (red) in order to initialize the cat state $\mathcal{C}_2^+$ with high fidelity in time $T = 0.3/K$.

and $\phi_0 = \hbar/2e$ is the flux quanta while $\Phi(t) = \Phi + \delta\Phi(t)$ is the classical flux through the SQUID loop. In our simulations, this flux is modulated around $\Phi = 0.2\phi_0$ at twice the resonator frequency $\omega_r$, $\delta\Phi(t) = \delta\Phi_0(t)\cos(2\omega_r t)$. In this expression, $\Phi_0(t)$ represents the slowly varying envelope of the modulation. As already discussed, for cat state initialization, $\Phi_0(t)$ is chosen to adiabatically change from zero to a maximum amplitude.

### Fourth-order expansion to map to Cassinian oscillator Hamiltonian

As usual, to map the above Hamiltonian to the Cassinian oscillator with time dependent two-photon drive, we expand the cosine term to the fourth order and make the rotating wave approximation. The resulting Hamiltonian is [7]

$$\hat{H} = \hbar\omega_r \hat{a}^\dagger \hat{a} - K\hat{a}^{\dagger 2}\hat{a}^2 + \mathcal{E}_p(\hat{a}^{\dagger 2}e^{-2i\omega_r t} + \hat{a}^2 e^{2i\omega_r t}),$$ (S8)

where we have defined

$$\omega_r = \omega_0 \sqrt{\cos\left(\frac{\Phi}{\phi_0}\right)\cos\left(\frac{\delta\Phi_0(t)}{\phi_0}\right)}, \quad \omega_0 = 4\sqrt{\frac{E_J E_C}{\hbar^2}}$$ (S9)

$$K = \frac{E_C}{2},$$ (S10)

$$\mathcal{E}_p = \frac{\sqrt{E_J E_C}}{2} \frac{\sin\left(\frac{\Phi}{\phi_0}\right)\sin\left(\frac{\delta\Phi_0(t)}{\phi_0}\right)}{\sqrt{\cos\left(\frac{\Phi}{\phi_0}\right)\cos\left(\frac{\delta\Phi_0(t)}{\phi_0}\right)}},$$ (S11)

with $E_C = e^2/C$ the charging energy. The strength of the two photon drive is governed by the amplitude of $\delta\Phi_0$. In practice, it cannot be made too large to avoid large change in the resonator frequency. As discussed in the main text, in order to encode an even parity cat state the resonator is initialized to vacuum state at $t = 0$, followed by an adiabatic increase in the two-photon drive amplitude. This is achieved by slowly increasing the amplitude of the flux modulation which, for simplicity, is here chose as $\delta\Phi_0(t) = \delta\Phi_0 \times t/\tau$.

### Exact simulation of the full Cosine potential

In order to account for higher-order effect or the rotating terms, we simulated the full Hamiltonian Eq. (S6) from $t = 0$ to $t = \tau$ with $\delta\Phi_0 = 0.04\phi_0$. The Wigner function of the resulting density matrix at $t = \tau$ is shown in

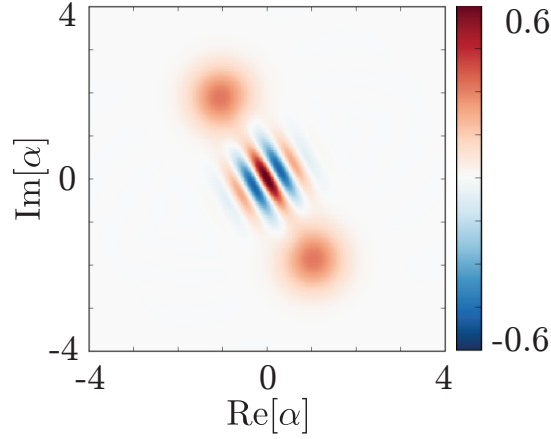

FIG. S4. The Wigner function of the cat state in the lab frame simulated using a realistic circuit of a resonator coupled to a flux-modulated SQUID. The full Cosine potential given by Eq. (S6) is simulated here.

Fig. S4. From Eq. (15) and (16), the size of the cat state is related to $\alpha = \sqrt{\mathcal{E}_p/K} \propto (E_J/E_C)^{1/4}$. As a result, initialization of large cat states requires large $E_J/E_C$. However, as this ratio is increased, higher-order terms become more important and can lead to reduction of the cat-state fidelity. Another consequence, as can be seen from Eq. (14), is that the frequency of the resonator $\propto \sqrt{E_J E_C}$ must be at least a few GHz to avoid thermal excitations from the bath. Keeping this in mind, we have used in our simulation the experimentally realistic parameters: $E_C/2\pi = 1.5$ MHz, $E_J/2\pi = 600$ GHz so that the estimated frequency $\omega_r/2\pi = 3.75$ GHz, nonlinearity $K/2\pi = 750$ KHz and two-photon drive strength $\mathcal{E}_p/2\pi = 3.7$ MHz. From simulations of the full Hamiltonian, we find that the cat state $|\mathcal{C}_\alpha^+\rangle$ with $|\alpha|^2 = 4.8$ is realized in time $t = \tau = 26.67$ $\mu$s with a fidelity of 99.4%. In other words, for these realistic parameters, the effect of higher-order terms appears to be minimal. Here and in the main paper, we have estimated fidelities using $F = \text{Tr}[\sqrt{\sqrt{\rho_{\text{target}}}\rho\sqrt{\rho_{\text{target}}}]$. We note that the rotation observed in Fig. S4 is due to the fact that this simulation was realized in the laboratory frame.

## VI. EFFECT OF SINGLE PHOTON DRIVE

In this section, we analyze a two-photon driven KNR with an additional single-photon drive. To simplify the analysis, we take $\kappa = 0$. The Hamiltonian is given by

$$\hat{H}_z = -K\hat{a}^\dagger\hat{a}^\dagger\hat{a}\hat{a} + \mathcal{E}_p(\hat{a}^{\dagger 2} + \hat{a}^2) + \mathcal{E}_z\hat{a}^\dagger + \mathcal{E}_z^*\hat{a}, \tag{S12}$$

where the phase of the single photon drive is defined with respect to the two-photon drive. Under a displacement transformation $D(\alpha_0)$ the Hamiltonian reads

$$\begin{aligned}
\hat{H}_z' &= [(-2K\alpha_0^2\alpha_0^* + 2\mathcal{E}_p\alpha_0^* + \mathcal{E}_z)\hat{a}^\dagger + \text{h.c.}] \\
&\quad + [(-K\alpha_0^2 + \mathcal{E}_p)\hat{a}^{\dagger 2} + \text{h.c.}] - 4K|\alpha|^2\hat{a}^\dagger\hat{a} - K\hat{a}^{\dagger 2}\hat{a}^2 - (2K\alpha_0\hat{a}^{\dagger 2}a + \text{h.c.}),
\end{aligned} \tag{S13}$$

where we have dropped the constant term $E = -K|\alpha_0|^4 + \mathcal{E}_p(\alpha_0^2 + \alpha_0^{*2}) + \mathcal{E}_z^*\alpha_0 + \mathcal{E}_z\alpha_0$ representing a shift in energy. For the coefficient of the $\hat{a}$, $\hat{a}^\dagger$ terms to vanish, we take

$$-2K\alpha_0^2\alpha_0^* + 2\mathcal{E}_p\alpha_0^* + \mathcal{E}_z = 0, \tag{S14}$$

such that

$$\hat{H}_z' = \left[\frac{-\mathcal{E}_z}{2\alpha_0^*}\hat{a}^{\dagger 2} + \text{h.c.}\right] - 4K|\alpha_0|^2\hat{a}^\dagger\hat{a} - K\hat{a}^{\dagger 2}\hat{a}^2 - 2K\alpha_0\hat{a}^{\dagger 2}a - (2K\alpha_0^*\hat{a}^\dagger\hat{a}^2 + \text{h.c.}). \tag{S15}$$

Following the derivation in the manuscript (Methods), $|0\rangle$ is an eigenstate of $\hat{H}_z'$ except for the first term which represents an off-resonant parametric drive of strength $|\mathcal{E}_z/2\alpha_0^*|$, detuned by $4K|\alpha_0|^2$. For $|\mathcal{E}_z/\alpha_0^*| \ll 4K|\alpha_0|^2$, fluctuations around $\alpha_0$ are small and $|0\rangle$ remains an eigenstate in the displaced frame. Again following the Methods section in the Manuscript, there are three solutions of Eq. (S18) which, for small $\mathcal{E}_z$, are of the form $-\alpha_0 + \epsilon$, $\epsilon$

and $\alpha_0 + \epsilon$ where $\alpha_0 = \sqrt{\mathcal{E}_p/K}$ and $\epsilon \sim \mathcal{E}_z/4\mathcal{E}_p$. Only two of these ($\alpha_0 + \epsilon$ and $-\alpha_0 + \epsilon$) satisfy the condition $|\mathcal{E}_z/\alpha_0| \ll 4K|\alpha_0|^2$. The large quantum fluctuations around the third solution makes it unstable. As a result, in the laboratory frame, the eigenstates of the system are $|\alpha_0 + \epsilon\rangle$ and $|-\alpha_0 + \epsilon\rangle$, where $\epsilon$ is a small correction ($\epsilon \to 0$ for $\mathcal{E}_z \ll 4\mathcal{E}_p$). In other words, the single-photon drive only slightly displaces the coherent components of the cat. From the above expression of the energy $E$, it is however clear that the degeneracy between the eigenstates $|\alpha_0 + \epsilon\rangle$ and $|-\alpha_0 + \epsilon\rangle$ is lifted by an amount $\delta_z = 4\mathrm{Re}[\mathcal{E}_z\alpha_0]$.

In the eigenspace spanned by $|\pm\alpha_0\rangle$, the single-photon drive can be written as $\delta_z\bar{\sigma}_z/2 + 2\mathrm{Im}[\mathcal{E}_z\alpha_0^*]e^{-2|\alpha_0|^2}\bar{\sigma}_x$. If $\mathcal{E}_z$ and $\mathcal{E}_p$ are real so that $\alpha_0$ is real, then $\mathrm{Im}[\mathcal{E}_z\alpha_0^*] = 0$ and hence the only effect of the single-photon drive is to lift the degeneracy between $|\pm\alpha_0\rangle$ or the logical $|\bar{0}\rangle$ and $|\bar{1}\rangle$.

## VII. EFFECT OF DETUNING

In this section we analyze the effect of detuning the two-photon drive from the resonator. The Hamiltonian is given by

$$\hat{H}_z = \delta\hat{a}^\dagger\hat{a} - K\hat{a}^\dagger\hat{a}^\dagger\hat{a}\hat{a} + \mathcal{E}_p(\hat{a}^{\dagger 2} + \hat{a}^2). \tag{S16}$$

Under a displacement transformation $D(\alpha_0)$ the new Hamiltonian reads

$$\hat{H}_z' = (-2K\alpha_0^2\alpha_0^* + 2\mathcal{E}_p\alpha_0^* + \delta\alpha_0)\hat{a}^\dagger + \text{c.c.} \\ + (-K\alpha_0^2 + \mathcal{E}_p)\hat{a}^{\dagger 2} + \text{c.c.} - 4K|\alpha_0|^2\hat{a}^\dagger\hat{a} + \delta\hat{a}^\dagger\hat{a} - K\hat{a}^{\dagger 2}\hat{a}^2 - (2K\alpha_0\hat{a}^{\dagger 2}a + \text{c.c.}), \tag{S17}$$

where we have dropped the constant term $E = -K|\alpha_0|^4 + \mathcal{E}_p(\alpha_0^2 + \alpha_0^{*2}) + \delta|\alpha_0|^2$ representing a shift in energy.

For the coefficient of the $\hat{a}, \hat{a}^\dagger$ terms to vanish,

$$-2K\alpha_0^2\alpha_0^* + 2\mathcal{E}_p\alpha_0^* + \delta\alpha_0 = 0, \tag{S18}$$

so that,

$$\hat{H}_z' = \frac{\delta\alpha_0}{2\alpha_0^*}\hat{a}^{\dagger 2} + \text{c.c.} - 4K|\alpha_0|^2\hat{a}^\dagger\hat{a} + \delta\hat{a}^\dagger\hat{a} - K\hat{a}^{\dagger 2}\hat{a}^2 - 2K\alpha_0\hat{a}^{\dagger 2}a - (2K\alpha_0^*\hat{a}^\dagger\hat{a}^2 + c.c.). \tag{S19}$$

Again, we follow the derivation outlined in the previous section to find that, if $|\delta| \ll 2\mathcal{E}_p$, then the eigenstates of the system are $|\pm\alpha_0\rangle$ where $\alpha_0 = \sqrt{(2\mathcal{E}_p + \delta)/2K}$. Because of the non-orthogonality of these states, the term $\hat{a}^\dagger\hat{a}$ has non-zero matrix elements $\langle\alpha_0|\hat{a}^\dagger\hat{a}| -\alpha_0\rangle = -|\alpha_0|^2e^{-2|\alpha_0|^2}$.

## VIII. EVOLUTION DURING THE GATE OPERATIONS

In this section, we provide more details on the performance of the single qubit $\hat{R}_{\bar{z}}(\theta)$, $\hat{R}_{\bar{x}}(\theta)$ and two-qubit gate. Fig. S5a) shows the probability for the system to be in the state $|\mathcal{C}_{\alpha_0}^-\rangle$ under evolution of the system with $H_0 + \hat{H}_z$ and single-photon loss for the system initially in $|\mathcal{C}_{\alpha_0}^+\rangle$. As expected, the probability shows a period of $\pi/4\mathcal{E}_z\alpha_0$. In the same way, Fig. S5b) shows the probability for the system to be in the $|-\alpha_0\rangle$ state under evolution of the system with $H_0 + \hat{H}_x$ and single-photon loss for the system initially in $|\alpha_0\rangle$. Again as expected, the probability shows a period of $\pi e^{2|\alpha_0|^2}/4\delta_x|\alpha_0|^2$.

As we saw in sections VI and VI, the mapping of the eigenstates to the coherent states $|\pm\alpha_0\rangle$ is valid only when $\mathcal{E}_z \ll 4K|\alpha_0|^3$ and $\delta \ll 2\mathcal{E}_p$. As a result, the gate performance depends on $\mathcal{E}_z/4K|\alpha_0|^3$ and $\delta/\mathcal{E}_p$. To demonstrate this, Fig. S5c) shows the probability of the system, initialized to $|\mathcal{C}_{\alpha_0}^+\rangle$ at $t = 0$, to be in the state $|\mathcal{C}_{\alpha_0}^-\rangle$ after a time $T = \pi/4\mathcal{E}_z\alpha_0$. The strength of the two-photon drive is fixed to $\mathcal{E}_p = 4K$ and to take into account the errors induced only due to large $\mathcal{E}_z$ we use $\kappa = 0$. Similarly, Fig. S5d) shows the probability of the system, initialized to $|\alpha_0\rangle$ at $t = 0$, to be in the state $|-\alpha_0\rangle$ after a time $T = \pi/4\delta|\alpha_0|^2\exp(-2|\alpha_0|^2)$, with $\mathcal{E}_p = K$ and $\kappa = 0$. As the figures indicate, with increasing $\mathcal{E}_z$ and $\delta$ the probability of achieving a perfect $Z$ and $X$ rotation decreases respectively. The small oscillations in the probability show that the eigenstates of the Hamiltonian are no longer coherent states.

Finally, for the entangling gate the system initialized to the product state $|\mathcal{C}_{\alpha_0}^+\rangle \otimes |\mathcal{C}_{\alpha_0}^+\rangle$. Figure S5(e) shows the time evolution of the probability for the system to be in the entangled state $|\psi_e\rangle = |\alpha_0, \alpha_0\rangle + i|\alpha_0, -\alpha_0\rangle + i|-\alpha_0, \alpha_0\rangle + |-\alpha_0, -\alpha_0\rangle$ and the expected periodicity, $\pi/8|\mathcal{E}_{zz}\alpha_0^2|$, is observed.

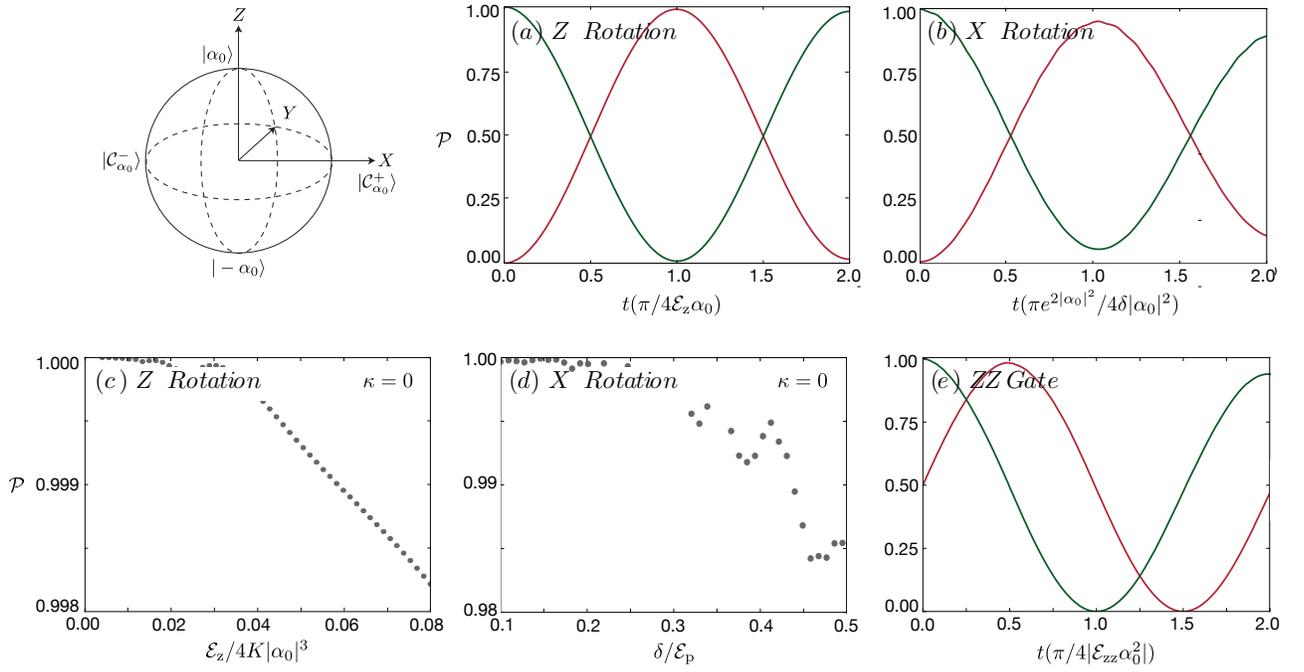

FIG. S5. (a) Probability for the system to be in state $|\mathcal{C}^+_{\alpha_0}\rangle$ (green) and $|\mathcal{C}^-_{\alpha_0}\rangle$ (red) for $\mathcal{E}_\mathrm{p} = 4K$, $\mathcal{E}_z = 0.8K$, $\alpha_0 = 2$. (b) Probability for the system to be in state $|\alpha_0\rangle$ (green) and $|-\alpha_0\rangle$ (red) for $\mathcal{E}_\mathrm{p} = K$, $\delta = K/3$, $\alpha_0 = 1$. Single photon loss for the simulations (a,b) is $\kappa = K/250$. (c) Probability for the system to be in the state $|\mathcal{C}^-_{\alpha_0}\rangle$ at time $T = \pi/4\mathcal{E}_z\alpha_0$ when it was initialized to the state $|\mathcal{C}^+_{\alpha_0}\rangle$ at $t = 0$. (d) Probability for the system to be in the state $|\alpha_0\rangle$ at time $T = \pi/4\delta|\alpha_0|^2\exp(-2|\alpha_0|^2)$ when it was initialized to the state $|-\alpha_0\rangle$ at $t = 0$. (e) Probability for the system to be in state $|\alpha, \alpha\rangle + |\alpha, -\alpha\rangle + |-\alpha, \alpha\rangle + |-\alpha, -\alpha\rangle$ (green) and $|\alpha, \alpha\rangle + i|\alpha, -\alpha\rangle + i|-\alpha, \alpha\rangle + |-\alpha, -\alpha\rangle$ (red) for $\mathcal{E}_\mathrm{p} = 4K$, $\mathcal{E}_{zz} = K/5$, $\alpha_0 = 2$. Single photon loss for the simulations is $\kappa = K/250$. The Bloch sphere is shown on the top left to illustrate the rotation axis.

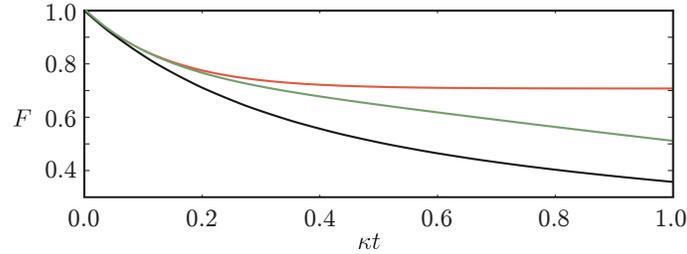

FIG. S6. Time dependence of the fidelity of the cat state, with $\kappa \neq 0$, $K = 0$, $\mathcal{E}_\mathrm{p} = 0$ (solid green line), $K = 20\kappa$, $\mathcal{E}_\mathrm{p} = 0$ (solid black line), and $K = 20$, $\mathcal{E}_\mathrm{p} = 4K$ (solid red line).

## IX. FIDELITY OF A CAT STATE IN KNR WITH AND WITHOUT TWO-PHOTON DRIVE

As explained in the main text, the fidelity of a cat state decreases faster in a KNR compared to a linear one. This is illustrated in Fig. S6 which compares the time-decay of the fidelity of a cat state $|\mathcal{C}^+_2\rangle$ initialized in a linear resonator (solid green line) and KNR (solid black line) with single-photon loss. In order to account only for non-deterministic errors, the fidelity in the lossy KNR is defined with respect to that of a lossless KNR. The cat state can be stabilized against Kerr-induced rotation and dephasing by the application of a two-photon drive chosen such that its amplitude $\mathcal{E}_\mathrm{p}$ satisfies Eq. (??). This is confirmed by the time dependence of fidelity in Fig. S6 (solid red line). As explained above, in a two-photon driven Kerr-resonator, a single-photon loss will only cause random jumps between the cat state $|\mathcal{C}^\pm_\alpha\rangle$ i.e., there is no energy relaxation, resulting in higher state fidelity than a linear cavity.

## X. QCMAP GATE WITH TWO-PHOTON DRIVING

Following Refs. [8, 9], we briefly describe the qcMAP gate protocol used to generate Fig. 4 of the main paper. The qubit is initialized to $(|e\rangle + |g\rangle)/\sqrt{2}$ and the resonator to the coherent state $|i\alpha_0\rangle$ (for simplicity we assume that $\alpha_0$ is a real number). The qubit and the resonator interact for a time $T_{\text{gate}}$ according to the ideal dispersive interaction $\hat{H}_{\text{disp.}} = \frac{g^2}{\Delta}\hat{a}^\dagger\hat{a}\sigma_z$, the full Jaynes-Cummings interaction, $\hat{H}_{\text{j.c.}} = \Delta\hat{\sigma}_z + g(\hat{a}^\dagger\hat{\sigma}_- + \hat{a}\hat{\sigma}_+)$ or the full JC interaction with two photon drive $\hat{H}_c = \hat{H}_{\text{j.c.}} - (\mathcal{E}_p\hat{a}^{\dagger 2} + \mathcal{E}_p^*\hat{a}^2)$. During this first step, the qubit and resonator ideally evolve to the entangled state $(|\alpha_0, g\rangle + |-\alpha_0, e\rangle)$ ignoring normalization. Next, an ideal displacement operation $D(\alpha_0)$ transforms the state to $(|2\alpha_0, g\rangle + |0, e\rangle)$. This is followed by an ideal qubit rotation conditioned on the number of photons in the resonator which is applied to disentangle the qubit from the resonator. This leaves the system in $(|2\alpha_0\rangle + |0\rangle) \otimes |g\rangle$. Finally, an ideal displacement of the resonator by $-\alpha_0$ results in the cat state $|\alpha_0\rangle + |-\alpha_0\rangle$. The Wigner function of this final state is shown in Fig. 4 of the main paper. In the simulations, we used the parameters: $g/2\pi = 111.4$ MHz, $\Delta/2\pi = 1.59$ GHz, $\kappa/2\pi = 7$ KHz, $\mathcal{E}_p = \mathcal{E}_p^\circ e^{i\phi}$, $\mathcal{E}_p^\circ/2\pi = 557$ KHz and $\phi = \pi/2$.

In order get an understanding for the phase and amplitude of the required two-photon drive, we examine $\hat{H}_c$ by expanding the Jaynes-Cummings interaction to the fourth order [10]

$$\hat{\hat{H}}_c = \Delta\hat{\sigma}_z + \frac{g^2}{\Delta}\hat{a}^\dagger\hat{a}\hat{\sigma}_z - \frac{g^4}{\Delta^3}(\hat{a}^\dagger\hat{a})^2\hat{\sigma}_z - (\mathcal{E}_p^*\hat{a}^2 + \mathcal{E}_p\hat{a}^{\dagger 2})$$

$$\sim \Delta\hat{\sigma}_z + \frac{g^2}{\Delta}\hat{a}^\dagger\hat{a}\hat{\sigma}_z - \frac{g^4}{\Delta^3}\hat{a}^{\dagger 2}\hat{a}^2\hat{\sigma}_z - (\mathcal{E}_p^*\hat{a}^2 + \mathcal{E}_p\hat{a}^{\dagger 2}), \tag{S20}$$

where we have assumed $g^2/\Delta - g^4/\Delta^3 \sim g^2/\Delta$.

To simplify the discussion, we now replace $\hat{\sigma}_z$ by its average value in the above expressions. In other words, we consider an infinite $T_1$ qubit. Going to a rotating frame, the above Hamiltonian then takes the form

$$\hat{\hat{H}}_c = -\frac{g^4}{\Delta^3}\hat{a}^{\dagger 2}\hat{a}^2\langle\hat{\sigma}_z\rangle - (\mathcal{E}_p^*\hat{a}^2 e^{i(g^2/\Delta)\langle\hat{\sigma}_z\rangle t} + \mathcal{E}_p\hat{a}^{\dagger 2}e^{-i(g^2/\Delta)\langle\hat{\sigma}_z\rangle t}). \tag{S21}$$

We are interested in time $T_{\text{gate}}$ such that the coherent states have rotated by $\pm\pi/2$ depending on the state of the qubit, i.e., $(g^2/\Delta)T_{\text{gate}} = \pi/2$. At this particular time, the above Hamiltonian reads

$$\hat{\hat{H}}_c = -K\hat{a}^{\dagger 2}\hat{a}^2\langle\hat{\sigma}_z\rangle - (\mathcal{E}_p^*\hat{a}^2 e^{i\pi\langle\hat{\sigma}_z\rangle/2} + \mathcal{E}_p\hat{a}^{\dagger 2}e^{-i\pi\langle\hat{\sigma}_z\rangle/2}) \tag{S22}$$

where $K = g^4/\Delta^3$ and $\langle\hat{\sigma}_z\rangle = \pm 1$. By comparing the above Hamiltonian with that of Eq. (1) of the main paper, we find that a two photon drive of amplitude $\mathcal{E}_p = -iK\alpha_0^2$ will ensure that the coherent states $|\pm\alpha_0\rangle$ or $(|\alpha_0, g\rangle + |-\alpha_0, e\rangle$ with the qubits) are the eigenstates of the Hamiltonian. In practice, the amplitude of $|\mathcal{E}_p|$ in the numerical simulations is slightly smaller than that predicted here because of the higher-order contributions of the Jaynes-Cummings Hamiltonian.

---

[1] Johansson, J., Nation, P. & Nori, F. Qutip: An open-source python framework for the dynamics of open quantum systems. *Computer Physics Communications* **183**, 1760–1772 (2012).

[2] Johansson, J., Nation, P. & Nori, F. Qutip 2: A python framework for the dynamics of open quantum systems. *Computer Physics Communications* **184**, 1234–1240 (2013).

[3] Wolinsky, M. & Carmichael, H. Quantum noise in the parametric oscillator: from squeezed states to coherent-state superpositions. *Physical review letters* **60**, 1836 (1988).

[4] Boutin, S., Andersen, C. K., Venkatraman, J., Ferris, A. J. & Blais, A. Resonator reset in circuit qed by optimal control for large open quantum systems. *arXiv preprint arXiv:1609.03170* (2016).

[5] Bourassa, J., Beaudoin, F., Gambetta, J. M. & Blais, A. Josephson-junction-embedded transmission-line resonators: From kerr medium to in-line transmon. *Physical Review A* **86**, 013814 (2012).

[6] Nigg, S. E. *et al.* Black-box superconducting circuit quantization. *Phys. Rev. Lett.* **108**, 240502 (2012). URL http://link.aps.org/doi/10.1103/PhysRevLett.108.240502.

[7] Wustmann, W. & Shumeiko, V. Parametric resonance in tunable superconducting cavities. *Physical Review B* **87**, 184501 (2013).

[8] Leghtas, Z. *et al.* Deterministic protocol for mapping a qubit to coherent state superpositions in a cavity. *Physical Review A* **87**, 042315 (2013).

[9] Vlastakis, B. *et al.* Deterministically encoding quantum information using 100-photon schrödinger cat states. *Science* **342**, 607–610 (2013).

[10] Boissonneault, M., Gambetta, J. M. & Blais, A. Dispersive regime of circuit qed: Photon-dependent qubit dephasing and relaxation rates. *Physical Review A* **79**, 013819 (2009).



\end{document}